\documentclass[12pt,preprint]{aastex}
\usepackage{multirow}

\shorttitle{M31 Velocity Vector III}
\shortauthors{van der Marel et al.}

\newcommand{\etal}{{et al.~}}

\newcommand{\kms}{\>{\rm km}\,{\rm s}^{-1}}

\newcommand{\Gyr}{\>{\rm Gyr}}
\newcommand{\kpc}{\>{\rm kpc}}
\newcommand{\Mpc}{\>{\rm Mpc}}
\newcommand{\Msun}{\>{\rm M_{\odot}}}

\begin{document}

\title{The M31 Velocity Vector.\\ 
III. Future Milky Way-M31-M33\\ 
Orbital Evolution, Merging, and Fate of the Sun}

\author{Roeland P.~van der Marel}
\affil{Space Telescope Science Institute, 3700 San Martin Drive, 
       Baltimore, MD 21218}

\author{Gurtina Besla}
\affil{Department of Astronomy, Columbia University, New York, NY 10027}

\author{T.J.~Cox}
\affil{Carnegie Observatories, 813 Santa Barbara Street, Pasadena, CA 91101}

\author{Sangmo Tony Sohn, Jay Anderson}
\affil{Space Telescope Science Institute, 3700 San Martin Drive, 
       Baltimore, MD 21218}


\begin{abstract}
We study the future orbital evolution and merging of the Milky Way
(MW)-M31-M33 system, using a combination of collisionless $N$-body
simulations and semi-analytic orbit integrations. Monte-Carlo
simulations are used to explore the consequences of varying all
relevant initial phase-space and mass parameters within their
observational uncertainties. The observed M31 transverse velocity from
Papers~I and~II implies that the MW and M31 will merge $t =
5.86^{+1.61}_{-0.72} \Gyr$ from now. The first pericenter occurs at $t
= 3.87^{+0.42}_{-0.32} \Gyr$, at a pericenter distance $r =
31.0^{+38.0}_{-19.8} \kpc$. In 41\% of Monte-Carlo orbits M31 makes a
direct hit with the MW, defined here as a first-pericenter distance
less than 25 kpc. For the M31-M33 system, the first-pericenter time
and distance are $t = 0.85^{+0.18}_{-0.13} \Gyr$ and $r =
80.8^{+42.2}_{-31.7} \kpc$.  By the time M31 gets to its first
pericenter with the MW, M33 is close to its second pericenter with
M31. For the MW-M33 system, the first-pericenter time and distance are
$t = 3.70^{+0.74}_{-0.46} \Gyr$ and $r = 176.0^{+239.0}_{-136.9}
\kpc$. The most likely outcome is for the MW and M31 to merge first,
with M33 settling onto an orbit around them that may decay towards a
merger later. However, there is a 9\% probability that M33 makes a
direct hit with the MW at its first pericenter, {\it before} M31 gets
to or collides with the MW. Also, there is a 7\% probability that M33
gets ejected from the Local Group, temporarily or permanently.  The
radial mass profile of the MW-M31 merger remnant is significantly more
extended than the original profiles of either the MW or M31, and
suggests that the merger remnant will resemble an elliptical
galaxy. The Sun will most likely ($\sim 85$\% probability) end up at
larger radius from the center of the MW-M31 merger remnant than its
current distance from the MW center, possibly further than $50 \kpc$
($\sim 10$\% probability). There is a $\sim 20$\% probability that the
Sun will at some time in the next $10 \Gyr$ find itself moving through
M33 (within $10 \kpc$), but while dynamically still bound to the
MW-M31 merger remnant. The arrival and possible collision of M31 (and
possibly M33) with the MW is the next major cosmic event affecting the
environment of our Sun and solar system that can be predicted with
some certainty.
\end{abstract}




\section{Introduction}
\label{s:intro}

Our Milky Way (MW) galaxy resides in a small group of galaxies called
the Local Group (LG; e.g., van den Bergh 2000). The three most massive
galaxies in the LG are all spirals: the MW, the Andromeda galaxy
(M31), and the Triangulum galaxy (M33), with mass ratios of $\sim
10:10:1$ (e.g., Guo \etal 2010; van der Marel \etal 2012, hereafter
Paper~II). Together, these galaxies dominate the LG mass. M33 lies at
about the same distance as M31 ($0.8 \Mpc$), and these two galaxies
most likely form a bound pair (McConnachie \etal 2009, hereafter M09;
Paper~II), as do the MW and M31 (van der Marel \& Guhathakurta 2008).

The orbits and interactions of the MW, M31 and M33 have been examined
in several previous studies. For example, Dubinski \etal (1996,
hereafter D96) and Cox \& Loeb (2008, hereafter CL08) presented
$N$-body simulations of the future MW-M31 interaction. Loeb \etal
(2005) and M09 presented $N$-body simulations of the past M31-M33
interaction. Innanen \& Valtonen (1977) used a Newtonion few-body
approach to study the future orbits of all three galaxies. And Peebles
\etal (2011) used a cosmological few-body approach (action modeling)
to study the past orbits of all three galaxies.

What all previous work has had common is that the relative
three-dimensional motion between M31 and either of the other two
galaxies was treated as a free parameter. The line-of-sight velocities
of M31 and M33 have long been well-known, and the proper motion of M33
was recently measured through water masers (Brunthaler \etal 2005).
However, a proper motion measurement of M31 remained elusive. As a
result, studies addressed and categorized possible past histories and
future outcomes, but only for a limited subset of uncertain initial
conditions.

The present paper is the third and final paper in a series. In Sohn
\etal (2012; hereafter Paper I) we reported the very first
proper-motion measurements of M31 stars in three different fields
observed with the Hubble Space Telescope (HST). In Paper~II we
combined these measurements with other techniques, and with an updated
understanding of the solar motion in the MW, to determine the
three-dimensional velocity vector of the M31 center of mass (COM) in
the Galactocentric rest frame. We also presented a combined analysis
of the masses of the MW, M31 and M33, based on literature results
combined with a new application of the LG timing argument.

With the results from Paper~II, all relevant dynamical quantities for
the MW-M31-M33 system are known. The galaxy distances are known to
$\sim 4$\% (measured as a fraction of the $\sim 1 \Mpc$ LG radius),
and the positional uncertainties on the celestial sphere are
negligible. The line-of-sight velocities are known to better than 1\%,
and the transverse velocities are known to $\sim 13$\% (measured as a
fraction of the $\sim 200 \kms$ LG virial velocity). The galaxy masses
are known to $\sim 30$\%, and the radial mass profiles are reasonably
well understood from cosmological simulations (e.g., Klypin \etal
2011). Hence, the calculation of the future dynamical evolution of the
MW-M31-M33 system is now entirely deterministic.

The goal of the present paper is to determine the future dynamical
evolution of the MW-M31-M33 system using $N$-body simulations, and to
use semi-analytic orbit integrations to assess and quantify the
variation in outcomes that is allowed by the observational
uncertainties. This is the first study of this topic based on fully
observationally constrained initial conditions. It is also the first
study to include detailed models of M33 in calculations of the
$N$-body evolution of the MW-M31 system. This allows us to study
several unique features, including the possibility that M33 may
collide with the MW before M31 does (see
Section~\ref{ss:orbitclasses}), the possibility that M33 may end up
ejected from the LG (see Section~\ref{ss:orbitclasses}; also Innanen
\& Valtonen 1977), and the possibility that M33 may accrete tidal
debris from the MW-M31 interaction, potentially including the Sun (see
Section~\ref{ss:sunfate} below).

The most important result from Paper~II is that the velocity vector of
M31 is statistically consistent with a radial (head-on collision)
orbit towards the MW. The inferred Galactocentric tangential velocity
of M31 is $V_{\rm tan} = 17.0 \kms$, with 1$\sigma$ confidence region
$V_{\rm tan} \leq 34.3 \kms$. This significantly constrains the future
dynamical evolution compared to what was known in the
past. Cosmological arguments about tidal torques from the local
Universe had merely constrained the $V_{\rm tan}$ to be $\lesssim 200
\kms$ (e.g., Gott \& Thuan 1978; Raychaudhury \& Lynden-Bell 1989;
Peebles \etal 2001).

No previous study has used an M31 velocity vector that is fully
consistent with the currently available observational
constraints. CL08 focused on initial conditions that provide a current
M31 tangential velocity of $V_{\rm tan} \approx 132 \kms$ (their
figure~6). Peebles \etal (2011) proposed a model in which M31 has
$V_{\rm tan,M31} = 100 \kms$. D96 constructed models with low
tangential velocities, $V_{\rm tan} = 20$ and $26 \kms$, which are
consistent with our new observational constraints. However, they used
a radial velocity $V_{\rm rad} = -130 \kms$, similar to CL08's $V_{\rm
  rad} \approx -135 \kms$. Both values yield a faster approach than
the value $V_{\rm rad,M31} = -109.3 \pm 4.4 \kms$ that is implied by
our latest understanding of the solar motion in the MW (see Section~4
of Paper II).

The MW and M31 mass distributions used in past work are also not fully
consistent with the currently available observational constraints.
D96 constructed two sets of models, one in which $M_{\rm LG} = 1.6
\times 10^{12} \Msun$ and one in which $M_{\rm LG} = 5.2 \times
10^{12} \Msun$, whereas the true mass is likely between these extremes
(see Paper~II). Also, they adopted a ratio $M_{\rm M31}/M_{\rm MW} =
2$ in all their models, while the actual ratio is likely closer to
unity. For example, the LSR circular velocity $V_0 = 239 \pm 5$ km/s
(Section~4.1 of Paper II) is similar to the rotation velocity of HI
gas in M31 (e.g., Chemin \etal 2009; Corbelli \etal 2010). CL08 used
galaxy masses $M_{\rm MW} = 1.0 \times 10^{12} \Msun$ and $M_{\rm M31}
= 1.6 \times 10^{12} \Msun$, but then also added an intra-group medium
of $2.6 \times 10^{12} \Msun$ for consistency with the LG timing
mass. However, recent insights (summarized in Paper II) suggest that
the LG timing mass is an overestimate, consistent with cosmic scatter,
so there is no need to force the models to match this mass
exactly. The calculations we present here without an intra-group
medium are likely to be more accurate.

We do not include in our study the effects of the LMC and SMC, the
next two most massive galaxies in the LG (van den Bergh 2000; Grebel,
Gallagher, \& Harbeck 2003).  Their combined mass is similar to that
of M33, making up $\sim 10$\% of the MW mass. The orbit of these
galaxies is such that they are moving rapidly away from the MW and may
not return for many Gyr (Besla \etal 2007; Shattow \& Loeb
2009). Their motion is away from the direction towards M31, to which
they have never been close (Kallivayalil \etal 2009). Their impact on
the overall future dynamics of the MW-M31-M33 System is therefore
likely to be small. The same is true for other LG dwarf galaxies,
which have masses well below those of M33 and the combined
LMC/SMC. This includes, e.g., M32, another well-known M31 satellite.
Even though it is the next most luminous galaxy in the LG, its
luminosity is only about one-tenth of the luminosity of M33.

The outline of this paper is as follows.  Section~\ref{s:methodology}
discusses the computational methodologies used for the $N$-body
simulations and semi-analytic orbit integrations.
Section~\ref{s:canonical} uses $N$-body simulations to calculate the
future of the M31-MW-M33 system, using a canonical set of initial
conditions that fall roughly midway in the observationally allowed
parameter ranges. It discusses the structure of the galaxies as the
simulation proceeds, and the possible fate of the Sun.
Section~\ref{s:semicalc} uses semi-analytic orbit integrations in a
Monte-Carlo sense to assess how the orbital evolution of the
MW-M31-M33 system changes when the initial conditions are varied
within their observationally allowed ranges. It discusses
characteristic pericenter and merger times, the fate of M33 (including
whether it collides with the MW before M31 does, or whether it is
ejected from the LG), and constraints on the M33 orbit around
M31. Section~\ref{s:conc} summarizes the main
conclusions. Appendix~\ref{a:coulomb} discusses the choice of Coulomb
logarithm in the dynamical friction formula used in the semi-analytic
orbit integrations.

\section{Computational Methodology}
\label{s:methodology}

We are concerned in this paper with the future dynamical evolution of
the system composed of the MW, M31, and M33. We use two complementary
methods to study this evolution, namely $N$-body simulations and
semi-analytic orbit integrations. The $N$-body simulations allow us to
study the evolution of each galaxy in detail, but due to computational
restrictions, only a small set of possible initial conditions can be
explored. The semi-analytic orbit integrations allow us to study only
the approximate motion of the COM of each galaxy. However, due to the
speed of the semi-analytic method, it allows a full exploration of the
parameter space of initial conditions. In the present section we
describe the respective methodologies used for these calculations.

\subsection{$N$-body calculations}
\label{ss:Nmethod}

We ran collisionless $N$-body simulations of the MW-M31-M33 system,
including only stars and dark matter. The calculations were performed
with the Nbody smoothed particle hydrodynamics code, GADGET-3
(Springel 2005). Typical numbers of particles used for the simulations
are listed in Table~\ref{t:galparam}.

In each of the galaxies, the gas comprises only a small fraction of
the total galaxy mass. We therefore chose not to include the gaseous
components of the galaxies in the simulations. This allowed us to run
higher-resolution simulations with larger numbers of particles. This
choice means that the overall dynamics of the interaction can be
followed with accuracy, but that issues such as hydrodynamic effects,
gas response, formation of gaseous streams, and star formation are not
addressed here. It would not be difficult to include these effects in
future numerical studies. CL08 did include gas and star formation in
their simulations of the MW-M31 system. They found that the features
thus induced are similar to what is normally seen in numerical
simulations of spiral-galaxy mergers.

Initial conditions for the galaxies were set up in the Galactocentric
frame, defined in Section~4 of Paper~II, with the MW starting at rest
at the origin. The initial position and velocity vectors ${\vec r}$
and ${\vec v}$ for M31 and M33, as well as the total virial masses
$M_{\rm vir}$ for all galaxies, were chosen to be consistent with the
observational results derived and summarized in Paper~II. Different
combinations of values were explored to produce different orbital
configurations, as discussed in subsequent sections.

The orientations of the galaxies, the scale lengths of their disks and
bulges, and the bulge masses $M_b$, were all chosen based on
literature values, as summarized in Table~\ref{t:galparam}. Disks of
mass $M_d$ were set up with exponential profiles with scale length
$R_d$. Warping, especially significant for M33 (e.g., Corbelli \&
Schneider 1997), was ignored. Bulges with mass $M_b$ were set up with
$R^{1/4}$ profiles. The bulge effective radius was taken to be $R_b =
0.2 R_{\rm disk}$. M33 was taken to be bulgeless; its nuclear
component (Corbelli 2003) was ignored, since it contributes negligibly
to the overall galaxy mass. To each galaxy we also added a massive
central black hole of mass $M_{\rm BH}$.

For a given total galaxy virial mass $M_{\rm vir}$, the virial mass of
the dark halo was taken to be $M_{\rm vir,h} = M_{\rm vir} - M_d - M_b
- M_{\rm BH}$. The halo concentration $c_{\rm vir}$ for each galaxy
was chosen to be consistent with cosmological simulations (Neto \etal
2007; Klypin \etal 2011). The concentration $c_{\rm vir}$ is defined
as $r_{\rm vir}/r_s$, where $r_s$ is the scale radius of the Navarro,
Frenk \& White (1997; hereafter NFW) profile that approximates the
dark halo.

In the simulations, we represent the dark halo of each galaxy by a
Hernquist (1990) density profile, with total mass $M_H$ and scale
length $a$, with no adiabatic contraction. A Hernquist profile is
similar to an NFW profile, but it drops off more steeply at large
radii. This has the advantage that the total mass is finite (see
discussion in Springel \etal 2005), which is not the case for an NFW
profile. For a given $M_{\rm vir}$ and $c_{\rm vir}$ we choose $M_H$ and
$a$ so that the corresponding Hernquist and NFW profiles have the same
asymptotic density for $r \rightarrow 0$, and the same enclosed mass
within $r_{\rm 200}$. The relevant equations are presented in
Appendix~A of Paper~II, to which we refer the reader for a discussion
of the various density profiles, scale radii, and masses that are
often encountered in the literature on dark halos. For example, if
$c_{\rm vir} = 10$, then $a/r_s = 2.01$, and $M_H/M_{\rm vir} =
1.36$. The total mass of the Hernquist model is larger than $M_{\rm
  vir}$, with the excess corresponding to the mass that is contained
at radii outside $r_{\rm vir}$.

For given $M_{\rm vir}$ and $c_{\rm vir}$, the mass of the disk $M_d$
of each galaxy was chosen to optimize the fit to the observed
amplitude of the galaxy rotation curve. For the MW we choose $M_d$ to
produce a circular velocity $V_c \approx 239 \kms$ at the solar radius
(McMillan 2011), consistent with the value used in Paper~II to correct
the observed motions of M31 and M33 for the reflex motion of the Sun.
For M31 we choose $M_d$ to produce a maximum circular velocity $V_c
\approx 250 \kms$ (Corbelli \etal 2010), and for M33 we choose $M_d$
to produce a maximum circular velocity $V_c \approx 120 \kms$
(Corbelli \& Salucci 2000). 

The central black hole mass for each galaxy was taken to be $M_{\rm
BH} = 3.6 \times 10^{-6} M_H$, motivated by the average black hole
demographics of galaxies. This does not yield a particularly accurate
fit to the known BH masses for the three galaxies under study here
($M_{BH,MW} = (4.1 \pm 0.6) \times 10^6 \Msun$, $M_{BH,M31} =
1.5_{-0.3}^{+0.9} \times 10^8 \Msun$, and $M_{BH,M33} < 3 \times 10^3
\Msun$; e.g., Gultekin \etal 2009 and references therein). However,
this does not matter for the present application, since either way,
the black holes have too little mass to influence the overall galaxy
dynamics during the interactions. The black holes are included here
primarily for numerical purposes, since they conveniently trace the
COM of the mostly tightly bound particles of each galaxy. Initially
this is the same as the COM of the galaxy as a whole. However, this
ceases to be true once the more loosely bound material becomes
significantly disturbed. In the following when we discuss the
evolution of the COM position of a galaxy, we merely follow the
position of its central black hole. When we discuss the evolution of
the COM velocity of a galaxy, we actually calculate a weighted average
over the luminous particles near the black hole.

Our approach starts the $N$-body simulations at the present epoch,
with initial positions and orientations reproducing the current
conditions. This is similar to the approach of D96, except that their
adopted M31 distance $D = 700 \kpc$ is smaller than the currently
favored value D = 770 $\pm$ 40 kpc (see Paper~II). By contrast, CL08
started their MW-M31 simulations 5 Gyr ago. So their models were not
tailored to exactly reproduce the observed location and spin
orientation of M31 at the present time, and are correct only in a
generic sense.

While our calculations were performed in the Galactocentric $(X,Y,Z)$
frame, we sometimes use a rotated set of coordinates with the same
origin, $(X',Y',Z')$ as defined in Section 6.1 of Paper II, to display
the results of the orbit calculations. We refer to this coordinate
system as the ``trigalaxy coordinate system''. The $(X',Y')$ plane,
which we will refer to as the ``trigalaxy plane'', is defined as the
plane that contains all three of the galaxies MW, M31 and M33 at the
present epoch. The $X'$-axis points from the MW to M31. As discussed
in Paper II, all three galaxies start out in the $(X',Y')$ plane, with
velocity vectors that are close to this plane. This implies that the
orbital evolution of the entire MW-M31-M33 system happens close to the
trigalaxy plane, with the ``vertical'' $Z'$-component playing only a
secondary role. For this reason, we show many of the three-dimensional
orbits calculated in subsequent sections only in their two-dimensional
($X',Y')$ projection.

\subsection{Semi-analytic orbit integrations}
\label{ss:semianmethod}

As mentioned, we also developed a semi-analytic method for calculating
the future motions of the MW, M31 and M33 due to their mutual
gravitational interaction. In this approximate method, each galaxy is
represented by a fixed one-component gravitational potential,
corresponding to a Hernquist density profile. For given $M_{\rm vir}$
and $c_{\rm vir}$, the total mass $M_H$ and scale length $a$ are
calculated as described before.

We write equations of motion that describe the position and velocity
of the COM of each galaxy (i.e., equations for 18 total phase
coordinates, 6 each for 3 galaxies). To calculate the gravitational
attraction in the equations of motion correctly, one would need to
integrate over all particles in the galaxies. To simplify matters, we
assume that the acceleration felt by a galaxy is as though all of its
mass were concentrated at its COM. Thus, the gravitational
acceleration felt by galaxy $j$ due to the Hernquist potential of
galaxy $i$ at distance $r$ equals $-G M_i / (r+a_i)^2$. The forces
thus implemented are non-symmetric and non-conservative. We therefore
apply a small correction at every time-step to ensure that the net
acceleration of the total COM of the whole system remains zero. We
also apply a constant density softening of $2 \kpc$ at the center of
each galaxy to avoid unphysical divergences. This softening is not a
particularly significant additional simplification, given that the
disks and bulges which dominate the gravitational potential at these
radii are not explicitly represented by our one-component models
either.

To include dynamical friction, we use the well-known Chandrasekhar
formula (Binney \& Tremaine 1987). Here too, we assume that the drag
felt by a galaxy is as though all of its mass were concentrated at its
COM. The Chandrasekhar formula is formally valid only for an infinite
homogeneous medium of density $\rho$ and velocity distribution $f(v)$.
For a galaxy $j$ undergoing friction from galaxy $i$, we substitute
the local $\rho$ and $f(v)$ in galaxy $i$, evaluated at the position
of COM$_j$. As usual, we assume $f(v)$ to be a Gaussian of dispersion
$\sigma$. The variation $\sigma(r)$ with radius in each galaxy was
taken from Hernquist (1990). 

The dynamical friction is proportional to the Coulomb logarithm $\log
\Lambda = \log (b_{\rm max} / b_{\rm min})$, where $b_{\max}$ and
$b_{\rm min}$ are the maximum and minimum impact parameter
contributing to the friction. We choose an expression for the Coulomb
logarithm that uses and expands the parameterization proposed and
tested by Hashimoto \etal (2003). We calibrate this expression using a
set of new $N$-body simulations. Appendix~\ref{a:coulomb} discusses
the parameterization, the choice of parameters based on the new
calibration, and the accuracy of the resulting semi-analytic orbit
integrations.

\section{Canonical $N$-body Model}
\label{s:canonical}

\subsection{Initial Conditions}
\label{ss:canonini}

\begin{figure}
\epsscale{0.33} \hbox to
\hsize{\plotone{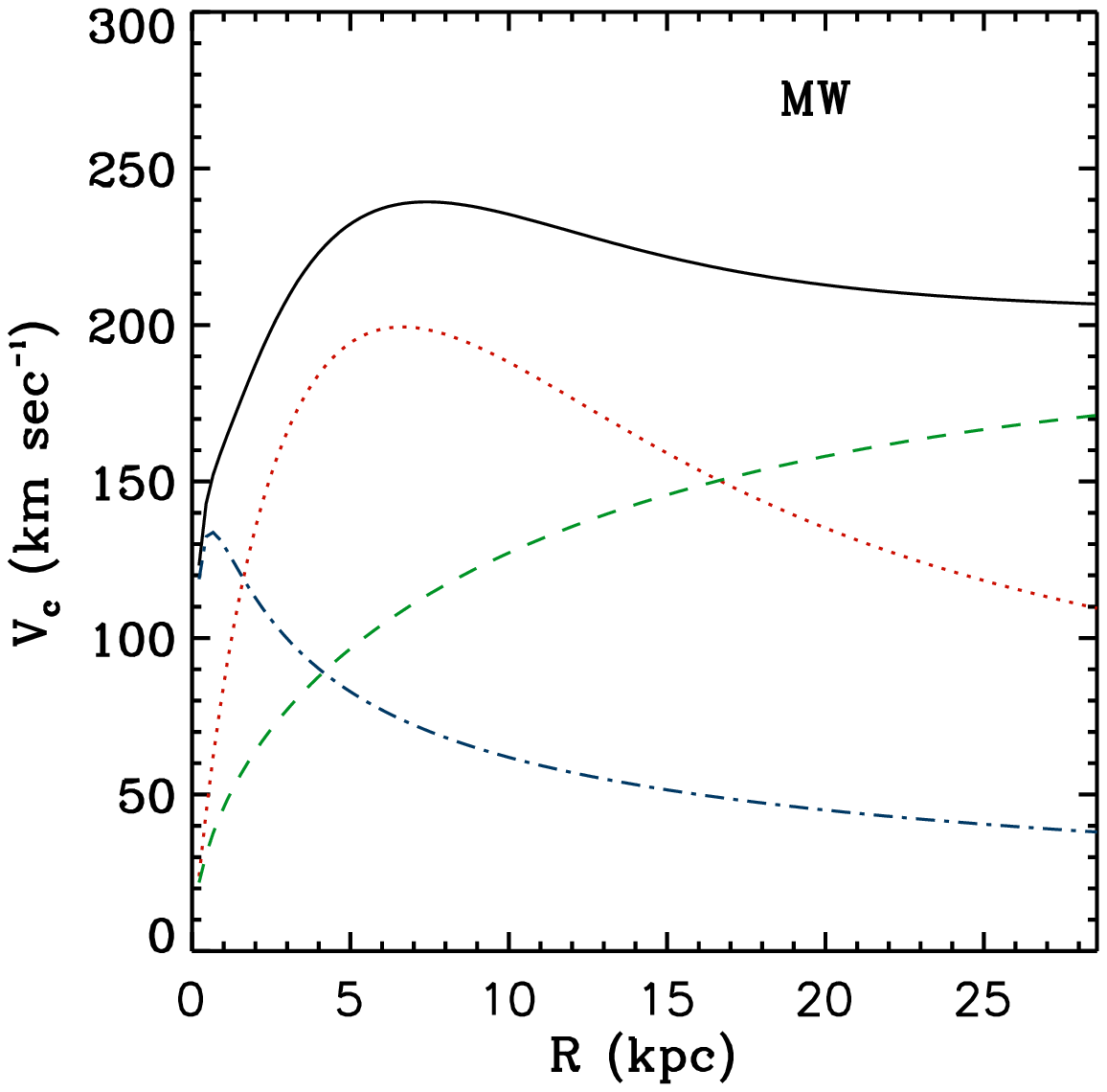}\hfill
  \plotone{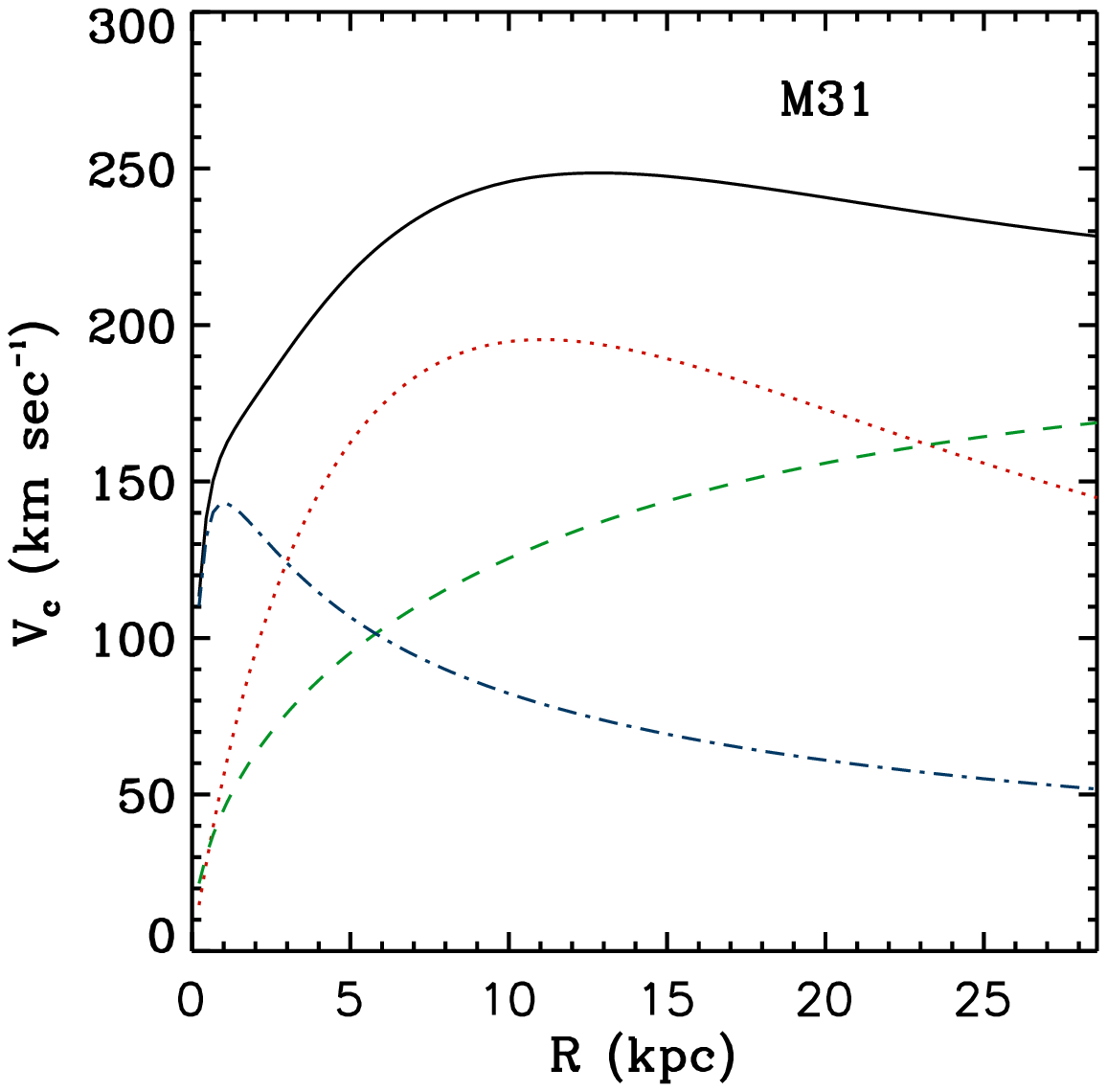}\hfill
  \plotone{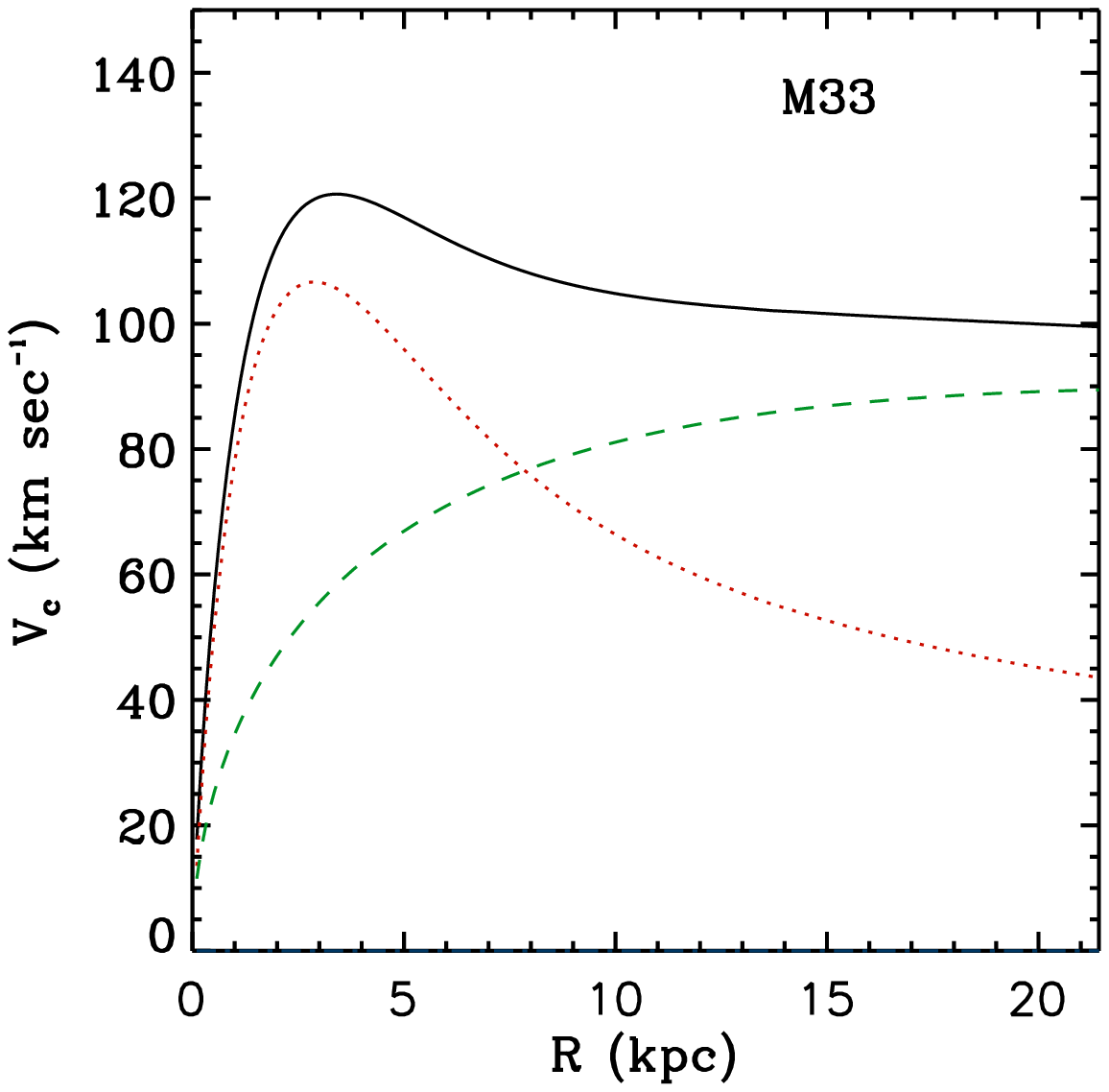}} 
\figcaption{Model rotation curves (black solid curves) of the galaxies
  MW, M31, and M33 (from left to right) at the start of the $N$-body
  simulations for the canonical model discussed in
  Section~\ref{s:canonical}. The individual contributions from the
  dark halo (green dashed ), disk (red dotted) and 
  bulge (blue dash-dotted) are indicated. The
  disk mass was chosen so as to reproduce approximately the observed
  maximum circular velocity for each galaxy: $V_c \approx 239 \kms$ at
  the solar radius for the MW (McMillan 2011), $V_c \approx 250 \kms$
  for M31 (Corbelli \etal 2010), and $V_c \approx 120 \kms$ for M33
  (Corbelli \& Salucci 2000).
\label{f:rotcurves}}
\end{figure}

We first study a canonical model for the MW-M31-MW system with main
characteristics summarized in Table~\ref{t:Norbits}. For this model we
adopt galaxy masses $M_{\rm MW,vir} = M_{\rm M31,vir} = 1.5 \times
10^{12} \Msun$ and $M_{\rm M33,vir} = 0.15 \times 10^{12} \Msun$.
These values are close to the midpoints of the observationally
constrained probability distributions derived in Paper~II. For a given
$M_{\rm vir}$, we calculate $c_{\rm vir}$ from the cosmological
simulation correlation presented by Klypin \etal (2011). The
corresponding Hernquist scale lengths, derived as described in
Section~\ref{ss:Nmethod} are $a_{\rm MW} = a_{\rm M31} = 62.5 \kpc$
and $a_{\rm M33} = 24.9 \kpc$. The disk masses adopted to produce the
desired maximum circular velocities are listed in
Table~\ref{t:Norbits}. The resulting rotation curves are shown in
Figure~\ref{f:rotcurves}.

The MW is initially at rest at the origin of the Galactocentric rest
frame. The adopted positions ${\vec r}_{\rm M31}$ and ${\vec r}_{\rm
M33}$ of M31 and M33 in the canonical model, respectively, are the
best estimates from Paper II (sections~4.2 and~6.1), based on the
known distances. The adopted velocity ${\vec v}_{\rm M33}$ of M33 is
the best estimate from Paper~II, based on the known line-of-sight
velocity and proper motion. The adopted velocity ${\vec v}_{\rm M31}$
of M31 is not exactly the best estimate from Paper II, but it agrees
with it to better than $7 \kms$ in each coordinate direction (about
1/3 of the observational error bars). The M31 line-of-sight velocity
in the canonical model is as observed, while the transverse motion is
$V_{\rm tan,M31} = 27.7 \kms$. For comparison the best estimate from
Paper~II is $V_{\rm tan,M31} = 17.0 \kms$, but all values $V_{\rm
tan,M31} \leq 34.3 \kms$ are consistent with the observational
constraints at 68.3\% confidence.\footnote{Originally the canonical
model pertained to our best estimate of the M31 transverse motion, but
that estimate in Paper II changed by a small amount at a late stage of
our project. Since the previously adopted value was still well within
the error bars, and the $N$-body simulations had already been run and 
analyzed, we decided not to redefine and recalculate a new
canonical model.}

\subsection{Angular Momentum} 
\label{ss:prograde}

The total orbital angular momentum of the MW-M31 system is ${\vec
L}_{\rm orb} \equiv \sum m_i {\vec r}_i \times {\vec v}_i$, where the
sum is over the COM properties of two galaxies.  If one ignores for
simplicity the roles of M33 and of dynamical friction, then this would
be a conserved quantity. In the Galactocentric rest frame, the MW is
initially at rest at the origin. Hence, ${\vec L}_{\rm orb}$ is simply
proportional to the cross product ${\vec r}_{\rm M31} \times {\vec
v}_{\rm M31}$ of the initial M31 position and velocity vectors. We
denote with ${\vec l}_{\rm orb}$ the unit vector in the direction of
${\vec L}_{\rm orb}$.

The unit vector in the direction of the MW spin angular momentum
equals ${\vec l}_{\rm sp,MW} = (0,0,-1)$ (the Sun rotates clockwise in
the $X,Y$ plane). The unit vector in the direction of the M31 spin
angular momentum can be calculated from the position and viewing
geometry of M31 (see Tables~\ref{t:galparam} and~\ref{t:Norbits}), and
equals ${\vec l}_{\rm sp,M31} = (-0.412,-0.767,-0.492)$.

An encounter between two galaxies is prograde for a galaxy if its spin
angular momentum is aligned with the orbital angular momentum, and it
is retrograde if the two are anti-aligned. Prograde encounters produce
more distortion and longer tidal tails than retrograde encounters
(e.g., Toomre \& Toomre 1972). The angle between the orbital and spin
angular momentum for galaxy $i$ equals $\beta_i \equiv \arccos ({\vec
  l}_{\rm orb} \cdot {\vec l}_{\rm sp,i})$. An angle $\beta =
90^{\circ}$ corresponds to a situation in which the orbital plane and
the galaxy plane are perpendicular. Smaller angles correspond to
prograde encounters, and larger angles to retrograde encounters. An
angle $\beta = 0^{\circ}$ corresponds to an in-plane prograde
encounter, and an angle $\beta = 180^{\circ}$ corresponds an in-plane
retrograde encounter.\footnote{Note that these extreme values are not
  encountered in the present situation. They would require M31 to lie
  in the MW disk plane or vice versa, neither of which is the case.}

Table~\ref{t:Norbits} lists the $\beta$ angles for the canonical
model. They are $\beta_{\rm MW} = 32.6^{\circ}$ and $\beta_{\rm M31} =
76.2^{\circ}$. Therefore, both galaxies undergo a prograde encounter
in this model. However, the spin-orbit alignment is better for the MW.
For M31, the spin axis is almost perpendicular to the MW-M31 orbital
plane.

\subsection{Orbital Evolution}
\label{ss:canonevol}

\begin{figure}
\epsscale{1.0} 
\plotone{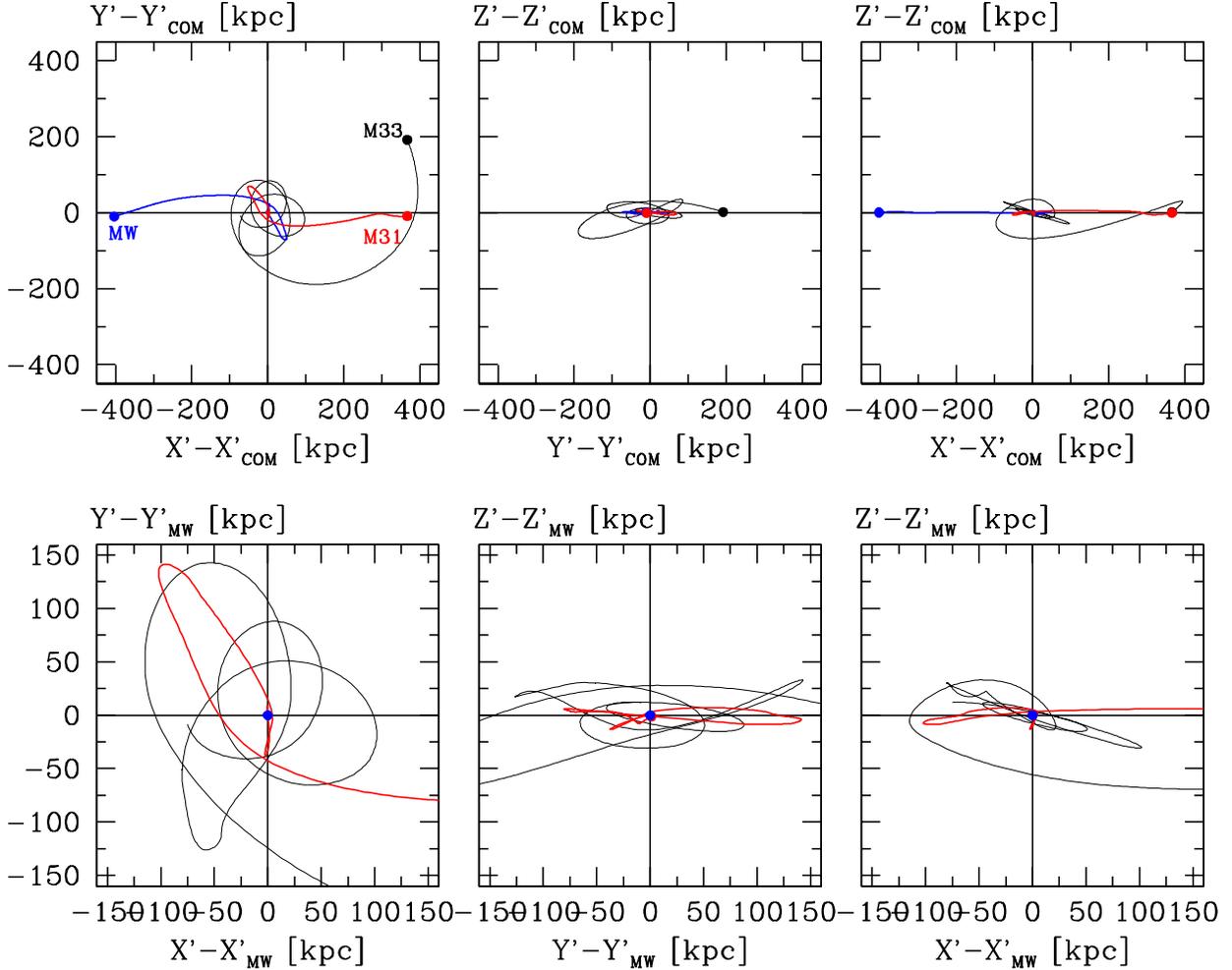} \figcaption{Orbital
evolution of the COM of the galaxies MW, M31, and M33, calculated with
$N$-body simulations for the canonical model discussed in
Section~\ref{s:canonical}. Each row shows three orthogonal
projections of the trigalaxy cartesian (X',Y',Z') system. The quantity
shown along the vertical axis is listed at the top left of each
panel. Top row: wide view fixed on the COM of the system. Bottom row:
central view fixed on the MW. The MW is shown in blue, M31 in red, and
M33 in black. Initial positions are shown with a dot. The MW and M31 merge
first. M33 settles into an elliptical, precessing, and
slowly-decaying orbit around them, in a plane that is close to the
M31-MW orbital plane.\label{f:canonorb}}
\end{figure}

\begin{figure}
\epsscale{0.60}
\plotone{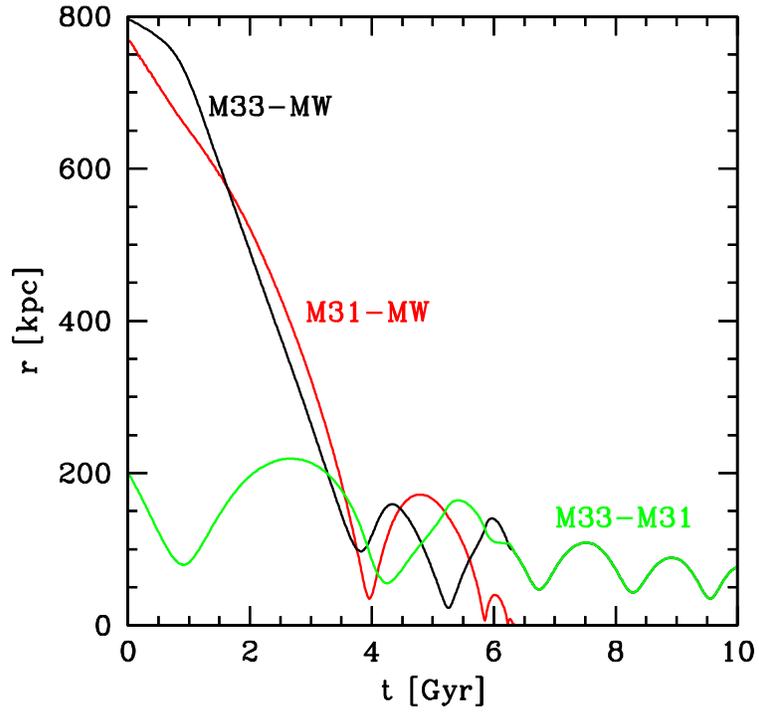} 
\figcaption{Galaxy separations in the MW-M31-M33 system as function of
time, calculated with $N$-body simulations for the canonical model
discussed in Section~\ref{s:canonical}. The M31-MW separation is shown
in red, the M33-MW separation in black, and the M33-M31 separation in
green.\label{f:canonsep}}
\end{figure}

\begin{figure}
\epsscale{0.60}
\plotone{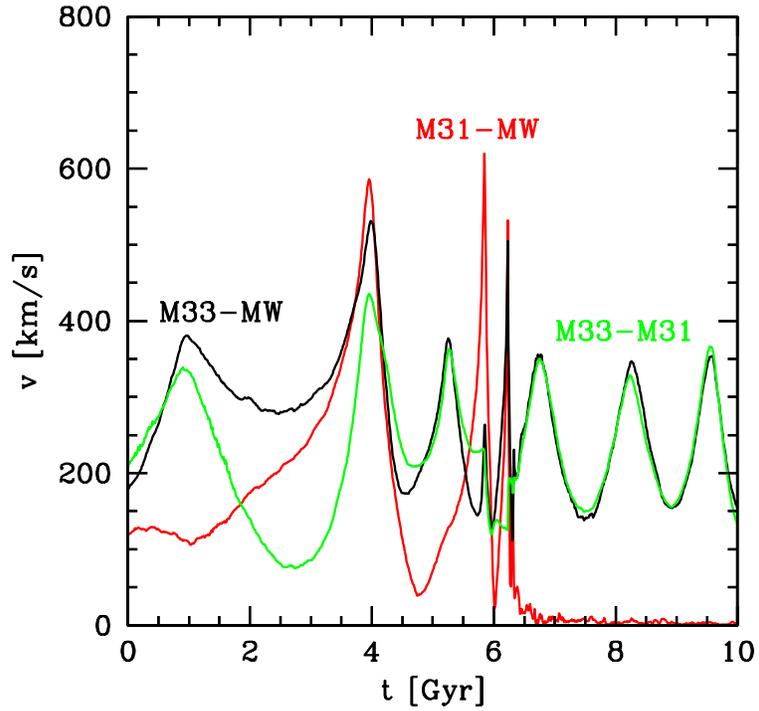} 
\figcaption{Relative velocities in the MW-M31-M33 system as function of
time, calculated with $N$-body simulations for the canonical model
discussed in Section~\ref{s:canonical}. The M31-MW relative velocity 
is shown in red, the M33-MW relative velocity in black, and the M33-M31 
relative velocity in green. Velocity maxima occur at orbital 
pericenters, and velocity minima at apocenters.\label{f:canonvel}}
\end{figure}

The orbital evolution for the canonical model, following the COM of
each galaxy, is shown in Figures~\ref{f:canonorb}, \ref{f:canonsep},
and~\ref{f:canonvel}. The top row of Figure~\ref{f:canonorb} shows
three orthogonal projections of the orbits in trigalaxy coordinates,
centered on the COM of the three-body system. The bottom row shows a
zoom-in in a frame that is comoving with the
MW. Figure~\ref{f:canonsep} shows the separations between M31-MW,
M33-MW and M33-M31, respectively, as function of time.
Figure~\ref{f:canonvel} shows the relative velocities as function of
time.

M31 has its next pericenter with the MW at $t = 3.97 \Gyr$ from now,
with a pericenter distance of $r = 35.0 \kpc$.  M31 then moves back
out to an apocenter distance of $r = 171.9 \kpc$ at $t=4.79 \Gyr$,
after which the orbit becomes almost directly radial towards the
MW. The two galaxies merge at $t = 6.29 \Gyr$.\footnote{For practical
  purposes in this paper, we consider two galaxies to have merged if
  their COM separation stays within $5 \kpc$ for an entire Gyr. The
  merging time is then defined as the latest time at which their
  separation exceeded $5 \kpc$.}

M33 has its next pericenter with M31 at $t = 0.92 \Gyr$, with a
pericenter distance of $r = 79.6 \kpc$. M33 then moves away from M31
to an apocenter distance of $r = 219.0 \kpc$ at $t = 2.66 \Gyr$. M33
reaches a first pericenter\footnote{As M33 approaches the MW (see
e.g. Figure~\ref{f:canonsep}), we refer to a minimum in the galaxy
separation as a ``pericenter'', even if originally M33 is not directly
orbiting the MW.} with respect to the MW of $r = 97.3 \kpc$ at $t =
3.83 \Gyr$. A second pericenter occurs much closer to the MW, $r =
23.0 \kpc$ at $t = 5.26 \Gyr$. After that, M33 settles into an
elliptical, precessing orbit around the M31-MW pair
(Figure~\ref{f:canonorb}, top left panel), in a plane that is close to
the M31-MW orbital plane (top middle and right panels). When the MW
and M31 merge, M33 is at a distance of $\sim 100\kpc$. The M33 orbit
decays slowly due to dynamical friction, which should eventually lead
to a merger with the M31-MW merger remnant. However, this will be many
Gyr later, since M33 is still in a relatively wide orbit (mean
distance $\sim 60\kpc$) at the time $t = 10 \Gyr$ when the simulation
was stopped.

The present-day relative velocity between M31 and the MW is $|{\vec
V}_{\rm M31}| = 110.6 \pm 7.8 \kms$ (Paper~II). However, the relative
velocity increases as M31 gets closer to the MW (see
Figure~\ref{f:canonvel}). At the first pericenter passage, the
relative velocity is as large as $586.0 \kms$. This explains why M31
and the MW subsequently recede to an apocenter distance as large as
$171.9 \kpc$. The relative velocity at subsequent pericenters remains
similarly high. However, the high velocities are maintained for
smaller periods of time during each subsequent pericenter, so that the
apocenter separations decrease with time.\footnote{The COM velocity
evolution in Figure~\ref{f:canonvel} depends on the exact way in which
the COM is defined. It is calculated here based on the luminous
particles near the central black hole. The velocity evolution of
loosely bound particles at large radii diverges substantially from
that of tightly bound particles near the galaxy center. This
decoupling is in fact one of the primary mechanisms for removing
orbital energy during the interaction (e.g., Barnes 1989).}

\subsection{Merger Process}
\label{ss:canonmerger}

Figure~\ref{f:canonsnaps} shows six snapshots (labeled a--f) of the
time evolution of the simulation, centered on the MW COM, and
projected onto the Galactocentric $(X,Y)$ plane (i.e., the MW disk
plane). Each panel spans $200 \times 200 \kpc$. At the start of the
simulation (the present epoch, $t=0$) M31 is at ${\vec r}_{\rm M31} =
(-378.9,612.7,-283.1) \kpc$ and M33 is at ${\vec r}_{\rm M33} =
(-476.1, 491.1, -412.9)$. So initially, both galaxies are located
outside the panels, off towards the top left. As the galaxies get
closer to the MW their orbits curve, making their approach directions
before pericenter almost parallel to the positive $Y$-axis. The past
orbital paths of the galaxies, as the simulation evolves, are
indicated with dashed lines. A full movie of the simulation ({\tt
figure\ref{f:canonsnaps}.mp4}) is distributed electronically as part of this
paper. The same projection and layout are used as in the panels of
Figure~\ref{f:canonsnaps}.

Panel (a) shows the MW disk after $t=3.00 \Gyr$, while it is still in
isolation, viewed pole-on. Particles that are color-coded red, and
which are followed throughout the simulation, are the ``candidate
suns'' discussed in Section~\ref{ss:sunfate} below. The initial MW
disk is not dynamically stable, and it develops a bar and spiral
features soon after the simulation starts. This is due to the
properties of the initial conditions, and in particular the somewhat
high disk mass needed to produce the observed circular velocity (we
use $M_d = 7.5 \times 10^{10} \Msun$, compared to, e.g., $M_d = 5
\times 10^{10} \Msun$ advocated by Klypin \etal 2002). The transient
disk features decrease as the disk secularly evolves through angular
momentum exchange and disk heating. We wished to minimize the impact
of the initial disk instability, given our interest to study the
future dynamical evolution of the candidate suns. We therefore evolved
the MW in isolation for $1 \Gyr$ before starting the actual
simulation, and did the same for M31 and M33.\footnote{Such evolution
in isolation was not done for the $N$-body simulations discussed in
Appendices~\ref{aa:firstM33} and~\ref{aa:retrograde} below. Those
simulations therefore still show the initial disk instabilities.}
Nonetheless, some secular evolution continues throughout the
simulation. The snapshot at $t=3.00 \Gyr$ therefore corresponds to a
slightly different equilibrium than the initial MW disk. It is in
principle possible to obtain a more stable initial axisymmetric disk
solution by changing the dark halo or bulge properties. However, the
MW does in fact posses a bar (e.g., Binney \etal 1997), so it is not
immediately obvious that this would produce a more realistic MW. We
therefore decided not to explore alternative models with more stable
disks.

Panel (b) shows the situation at the time $t = 3.97 \Gyr$ of the first
MW-M31 pericenter. The galaxies partially overlap at this time. The
galaxy spins are such that the encounter is prograde for both
galaxies. However, the relative velocity between the galaxies is
large, so that the immediate damage they inflict on one another is
limited. At this time, M33 has already moved to a position on the
negative $Y$ side of the MW, and is $0.14 \Gyr$ past its MW
pericenter. The past COM orbits of M31 and M33 are indicated with
dotted curves.

Panel (c) shows the situation at the time $t = 4.47 \Gyr$,
when M31 is getting close to its first apocenter with MW. Due to the
prograde nature of the encounter, stars are thrown out of both
galaxies along extended tidal tails. The tails are roughly
bisymmetric. The tails are more prominent for the MW, since its
encounter is more prograde than for M31 (see D96 for a detailed study
of the formation of tidal tails in the MW-M31 interaction, and how
their development depends on model parameters). M33 has just passed
its second pericenter with M31, the first one having happened when the
galaxies were still far from the MW (at $t = 0.92 \Gyr$). Also, M33
has just passed its first apocenter with the MW, and is now starting
to fall back onto it. Due to tidal perturbations from both M31 and the
MW, M33 has developed the pronounced S-shaped structure that is
characteristic of tidal stripping of a satellite (e.g., Odenkirchen
\etal 2001).

Panel (d) shows the situation at the time $t = 6.01 \Gyr$ when M31 has
passed its second pericenter with the MW, and is now at its second
apocenter.  The galaxies are separated now by only $40.0 \kpc$, and
the process of merging has started. At this time M33, is also
approximately at its second apocenter with respect to the MW. However,
its separation from the MW is still $r = 139.4 \kpc$.

Panel (e) shows the situation at the time $t=6.38 \Gyr$, approximately
$0.1 \Gyr$ after the two galaxies have merged. The structure of the
MW-M31 remnant is highly asymmetric and unrelaxed at this time. The MW
and M31 stars in the remnant have not yet mixed. The MW stars are
located on average at lower $X$ than the M31 stars. M33 continues to
orbit the MW-M31 remnant, and is at $r = 90.6 \kpc$ from it.

Panel (f) shows the situation at the time $t=10.00 \Gyr$, which is
when the simulation was stopped. The MW-M31 remnant has had several
Gyr to relax dynamically, and its structure has become fairly smooth
and symmetric, resembling an elliptical galaxy. The original MW and
M31 stars have mixed throughout the remnant. M33 still orbits the
remnant on an elliptical, precessing, and slowly decaying orbit. It is
at a distance $r = 75.0 \kpc$ from it. Tidally stripped stars from M33
contribute to the halo of the MW-M31 remnant. However, M33 is still
easily recognizable as a separate galaxy, and this has in fact
remained the case throughout the entire simulation.

\subsection{Remnant Structure}
\label{ss:remnant}

The three panels of Figure~\ref{f:rembygal} show the distribution of
luminous particles at the end of the simulation for the three
different galaxies. The MW and M31 particles have become mixed around
a common COM. However, the remnant is not yet fully relaxed, since
particles originating from the two different galaxies still have a
somewhat different spatial distribution. This is evident both from a
difference in ellipticity near the center, and from a difference in
the structure of shells and tails at large radii. 

The distribution of M33 stars at the end of the simulation is markedly
different from that of the MW and M31 stars. M33 still largely
maintains its own identity in a bound core.  However, 23.5\% of its
luminous particles have been tidally stripped and are now located
outside the M33 Roche radius ($17.6 \kpc$ at the end of the
simulation). These stripped particles occupy wrapped streams that
populate the halo of the MW-M31 merger remnant. These streams do not
lie along the location of the orbit, primarily due to a combination of
two effects. First, particles continue to be affected by M33's gravity
even after they have been stripped (Choi, Weinberg, \& Katz 2007). And
second, stripped stars have a velocity component out of the orbital
plane since the M33 disk is not aligned with that plane.

Comparison of Figure~\ref{f:canonsnaps}f to~\ref{f:canonsnaps}a shows
that the MW-M31 merger remnant is significantly more radially extended
than its progenitor galaxies. Figure~\ref{f:remnant} shows the
projected surface density profile of luminous MW and M31 particles in
the remnant at the end of the simulation ($t = 10 \Gyr$). The profile
roughly follows an $R^{1/4}$ profile at radii $R \gtrsim 1 \kpc$,
characteristic of elliptical galaxies. This is consistent with the
large literature supporting the assertion that roughly-equal mass
mergers of spiral galaxies form remnants that can be classified as
elliptical galaxies (e.g., Barnes 1998). However, what fraction of
ellipticals form this way remains an open question (e.g., Naab \&
Ostriker 2009, and references therein).

Both the MW and M31 are fairly typical spirals in our
simulations. There is therefore no reason to expect the properties of
the merger remnant to be very different from what has been found
generically for mergers of spiral galaxies. Indeed, CL08 studied the
merger remnant in their MW-M31 simulations in some detail, and found
its properties to be consistent with those of elliptical
galaxies. While their simulations differ from ours in a number of key
areas (see discussion in Section~\ref{s:intro}), there is no
particular reason to expect that this would change the {\it generic}
features of the remnant. For these reasons, we do not present a
detailed analysis of the merger remnant in our simulations. Such an
analysis might have included a characterization of triaxiality,
ellipticity, boxyness/diskyness, fundamental-plane position, rotation
rate, and deviations of velocity distributions from Gaussians (e.g.,
Naab \& Burkert 2003; Naab, Jesseit, \& Burkert 2006; CL08). These
characteristics can all be compared to what is observed for samples of
ellipticals. Nonetheless, even without such a study, it seems clear
that the CL08 conclusion that the MW-M31 merger remnant should
resemble an elliptical galaxy still holds.

\subsection{The Fate of the Sun}
\label{ss:sunfate}

As in CL08, it is of interest to address what the future fate of the
Sun might be as the MW-M31-M33 system evolves. We do this by
identifying ``candidate suns'' in the simulation, and by following
their fate as time progresses. We identify the candidate suns from
their phase space properties at time $t = 3 \Gyr$ in the simulation,
well before M31 and M33 arrive near the MW. Candidate suns are not
identified at the start of the simulation in order to minimize the
impact of transient initial features in the MW disk.

We probably do not understand the secular evolution of the MW well
enough to predict how the phase space coordinates of the real Sun will
change over the next $3 \Gyr$, so we neglect any such evolution. Also,
we cannot predict at what azimuth in the $(X,Y)$ plane the Sun will be
$3 \Gyr$ from now. The Sun is at $R_{\odot} \approx 8.29 \kpc$ from
the Galactic Center, and the circular velocity at this radius is
$V_{\odot} \approx 239 \kms$ (see discussion in Paper II). Hence, $3
\Gyr$ from now the Sun will have made $\sim 14$ orbital revolutions. A
$\pm 3.6$\% uncertainty in the circular velocity, which is fairly
realistic, therefore produces a $\pm \pi$ uncertainty in azimuth.
Based on these considerations, we identify as candidate suns those
particles in the simulated MW at $t=3\Gyr$ that: (a) are within $0.10
R_{\odot}$ from the circle $R = R_{\odot}$ in the equatorial plane;
(b) have an in-plane velocity $(V_X^2+V_Y^2)^{1/2}$ that agrees with
$V_c$ at $R = R_{\odot}$ to within $0.10 V_c$; and (c) have an
out-of-plane velocity that satisfies $|V_Z| < 30 \kms$. A total of
8786 luminous particles meet these criteria, i.e., 1.0\% of the
total. Our criterion for identification of candidate suns is more
strict than that of CL08, who selected particles at the solar radius
irrespective of velocity.

The candidate suns are shown in red in the simulation snapshots of
Figure~\ref{f:canonsnaps}, and also in the movies distributed with
this paper (but only for $t \geq 3\Gyr$). They start out as a ring of
particles in panel (a). But due to violent relaxation and phase mixing
they end up distributed throughout the merger remnant at the end of
the simulation in panel (f). Figure~\ref{f:remnant} shows the
projected surface density profile of the candidate suns in the MW-M31
merger remnant at the end of the simulation. The profile is somewhat
less centrally concentrated than the surface-density profile of all
particles in the remnant.  Hence, a candidate sun resides on average
at a larger radius in the remnant than an average particle. This is
because the Sun is initially not particularly tightly bound within the
MW. Since the initial and final binding energies of a particle in an
interaction are correlated, this also remains true at the end of the
simulation in the MW-M31 remnant.

Figure~\ref{f:sunhist} shows the radial distribution of candidate suns
with respect to the center of the MW-M31 remnant, at the end of the
$N$-body simulation ($t = 10 \Gyr$) for the canonical model. All the
candidate suns initially start out at $r \approx R_{\odot} \approx
8.29 \kpc$. At the end of the simulation, 14.6\% of the candidate suns
reside at $r< R_{\odot}$ and 85.4\% at $r>R_{\odot}$. A fraction
10.4\% reside at $r > 50 \kpc$ and a fraction 0.1\% at $r > 100
\kpc$. Therefore, our actual Sun will most likely migrate outward
during the merger process, compared to its current distance from the
Galactic Center (consistent with the earlier findings of CL08). There
is a small but significant probability that the Sun will migrate out
to very large radius. However, no candidate suns become entirely
unbound from the MW-M31 merger remnant in the simulation.

The radial distribution of candidate suns in Figure~\ref{f:sunhist}
pertains to a snapshot at a fixed time. While this distribution is
reasonably stable with time, this does not mean that each individual
candidate sun remains at a fixed radius. Each individual candidate sun
orbits the MW-M31 merger remnant, so its radial distance from the
center is time-dependent. This is true in particular for candidate
suns that move out to large distances, since those tend to be on
relatively radial orbits. So even if a candidate sun resides far from
the center of the merger remnant for most of the time, it may find
itself plunging rapidly through the central region at regular
intervals. The distributions of orbital pericenters, apocenters,
semi-major axis lengths, or time-averaged distances could in principle
be calculated for the solar candidates, and those would differ from
the distribution in Figure~\ref{f:sunhist}.

Some of the candidate suns migrate out far enough to overlap with the
range of radii where M33 orbits the MW (or the MW-M31 merger remnant)
after $t \approx 5 \Gyr$ (see Figure~\ref{f:canonsep}). It is
therefore theoretically possible that candidate suns could be accreted
by M33. However, M33 moves rapidly around the MW (see
Figure~\ref{f:canonvel}), and its orbit is very different from that of
typical stars in the MW, even those ejected into tidal tails.
Therefore, most candidate suns that pass close to M33 will undergo
flyby encounters, and will be not be accreted by M33 (i.e., become
bound). In the canonical model, none of the 8786 candidate suns became
bound to M33. This sets an upper limit of $\sim 10^{-4}$ to the
probability of the Sun ever becoming bound to M33.

While the probability of candidate suns becoming bound to M33 is
small, the probability is larger that a candidate sun might find
itself temporarily inside M33. For each candidate sun in the
simulation we studied whether it ever came within 10 kpc of the M33
COM (despite M33 always being more than $23 \kpc$ from the MW COM
throughout the simulation; see Figure~\ref{f:canonsep}). A total of
1762 candidate suns met this criterion over the $10 \Gyr$ of the
simulation. Therefore, the probability is 20.1\% that the Sun will at
some time in this future period find itself moving through M33,
although dynamically still being associated with the MW-M31
remnant.\footnote{The probability is smaller (13.9\%) when calculated
  over all MW particles, and not just candidate suns.}

\section{Semi-Analytic Orbit Calculations}
\label{s:semicalc}

\subsection{Initial Conditions}
\label{ss:semiini}

The initial conditions used for the canonical orbital scenario in
Section~\ref{s:canonical} form only one of many possibilities that are
consistent with the observational uncertainties on the galaxy masses
and phase-space coordinates derived in Paper~II. We therefore used a
Monte-Carlo scheme to create $N=1000$ initial conditions that explore
the full parameter space of possibilities. Table~\ref{t:orbits} lists
the observational and theoretical constraints on the galaxy masses,
halo concentrations, and initial phase-space coordinates that were
used to generate the initial conditions.

We studied the orbital evolution for each set of initial conditions,
and determined the probability $p$ with which certain orbital features
occur (e.g., a direct hit between galaxies). If a feature occurs in
$N_f$ orbits, then $p = N_f/N$ with random uncertainty $\Delta p
\approx \sqrt{N_f}/N$. For $N=1000$, $p=10$\% yields $\Delta p = 1$\%,
and $p=1$\% yields $\Delta p = 0.3$\%. For larger $N$, parameter space
would be explored more fully and the random uncertainties $\Delta p$
would be smaller. However, systematic uncertainties due to the
simplifying assumptions that underly the semi-analytic calculations
would stay the same. Since these likely dominate the random
uncertainties, we decided that $N=1000$ was sufficient.

The initial masses $M_{\rm vir}$ for the galaxies were drawn as in
Section~5 of Paper II. This combines observational constraints on the
masses of the individual galaxies, with the timing-argument
constraints on the total mass of the MW and M31 (including cosmic
scatter, following Li \& White 2008). At given $M_{\rm vir}$, we
calculate $c_{\rm vir}$ from the cosmological simulation correlation
presented by Klypin \etal (2011). We add a random scatter of $0.1$ dex
in $\log_{10} c_{\rm vir}$, consistent with the simulations of Neto
\etal (2007).

The initial phase-space coordinates of M31 and M33 were drawn as in
Sections~4.2 and~6.1 Paper~II, respectively, based on the observed
positions, distances, line-of-sight velocities, and transverse
velocities. Thus employed, this scheme propagates all observational
distance and velocity uncertainties and their correlations, including
those for the Sun\footnote{Uncertainties in the RA and DEC of M31 and
  M33 are negligible and were ignored.}.

For each set of initial conditions, we calculated the binding energies
of the MW-M31 and M33-M31 systems, respectively. The MW-M31 system was
found to be bound in all cases. This is as expected, since an unbound
chance encounter of two galaxies like the MW and M31 would be quite
unlikely given the local density of spiral galaxies (van den Bergh
1971). The M33-M31 system was found to be bound in 95.3\% of
cases. There is observational evidence from both HI (Braun \& Thilker
2004) and star count maps (M09) for tidal features indicative of past
interactions between these two galaxies. M09 have argued that M33 and
M31 {\it must} be bound for the observed tidal features to have
formed.  This is the approach that we will take here. So as in Section
6.3 of Paper~II we remove from consideration the small subset of
initial conditions in which M33 and M31 are not bound.

Figure~\ref{f:prograde} shows the distributions of $\beta_{\rm MW}$
and $\beta_{\rm M31}$ for the initial conditions, being the angles
between the galaxies initial spin and orbital angular momenta as
defined in Section~\ref{ss:prograde}. Both distributions are quite
broad. For the MW, 72.2\% of the orbits are prograde, and 27.8\% are
retrograde. For M31, 59.0\% of the orbits are prograde, and 41.0\% are
retrograde. So for both galaxies, a prograde encounter is the most
likely outcome, but not by a large factor. For M31, 41.4\% of orbits
have $|\beta_{\rm M31} - 90^{\circ}| < 30^{\circ}$, so that nearly
orthogonal encounters are quite probable. For the MW, this fraction is
only 28.0\%. Based on these considerations, the initial MW-M31
encounter will generally perturb the MW more strongly than M31, as was
the case in the canonical model (see Section~\ref{ss:canonmerger}).

The reason that both prograde and retrograde encounters are possible
is that the M31 velocity is directed almost radially towards the MW.
Hence, the orbital angular momentum is small, and its direction is
poorly determined. Specifically, one may consider the M31 transverse
velocity space $(v_W,v_N)$, in which one value $(v_W,v_N)_{\rm rad}$
corresponds to the velocity vector for a radial orbit (see Figure~3 of
Paper~II). Most of the velocities inside the observational error
ellipse lie on one side of $(v_W,v_N)_{\rm rad}$, and those yield a
prograde encounter for the MW. However, some of the velocities lie on
the other side of $(v_W,v_N)_{\rm rad}$, and those yield a retrograde
encounter for the MW.

\subsection{Distributions of Orbital Characteristics}
\label{ss:explore}

For each of set of Monte-Carlo generated initial conditions we
calculated the future orbital evolution of the MW-M31-M33 system with
the semi-analytic orbit-integration method of
Section~\ref{ss:semianmethod}. All orbits were integrated for a
sufficiently long time to be able to quantify and classify the future
evolution. For all orbits we determined the same characteristic
quantities (pericenters, apocenters, merging times, etc.) as discussed
in Section~\ref{ss:canonevol}. 

For all initial conditions, M33 is moving towards M31 at the present
time, albeit with a significant tangential component (see e.g. top
left panel of Figure~\ref{f:canonorb}, and Section~6.1 of Paper
II). M31 is moving towards the MW, and is pulling M33 with it. Each
pair of galaxies is therefore heading for a pericenter
passage. Figure~\ref{f:thist}a shows histograms of the times of these
pericenter passages, and Figure~\ref{f:thist}b shows histograms of the
corresponding pericenter distances. Figure~\ref{f:thist}c shows a
histogram of the times at which the MW and M31 merge.

For the MW-M31 system, the first-pericenter time and distance are $t =
3.87^{+0.42}_{-0.32} \Gyr$ and $r = 31.0^{+38.0}_{-19.8} \kpc$. Here
and henceforth, values quoted for a quantity refer to the median and
surrounding $68.3$\% confidence regions in the Monte-Carlo
distribution. In 41.0\% of orbits the MW-M31 first-pericenter distance
is less than 25 kpc. Taking into account the sizes of the galaxy
disks, we consider this arbitrary cutoff to denote a ``direct
hit''.\footnote{We use the same definition of ``direct hit''
throughout this paper, independent of which pair of galaxies has a
close passage. The disk exponential scale lengths of all three
galaxies are well below 25 kpc (see Table~\ref{t:galparam}). Hence, a
direct hit does not necessarily mean that the densest central parts of
the galaxies are colliding. However, all three galaxies have disks
that extend to many exponential scale lengths. For example, the stars
in the M31 disk can be traced to beyond 20 kpc (Courteau \etal 2011)
and the HI gas to some 40 kpc (Corbelli \etal 2010). In M33, the
smallest of the three galaxies, the stars in the disk can be traced to
at least 4 kpc (Kent 1987) and the HI gas to some 15 kpc (Corbelli \&
Schneider 1997).} This high fraction is no surprise, since the data
constrain the relative orbit to be close to radial. An example
$N$-body simulation of such a direct hit is discussed in
Appendix~\ref{aa:direct}. Its initial conditions are listed in
Table~\ref{t:Norbits} in the column labeled ``direct-hit''. A full movie
({\tt figure\ref{f:direct}c.mp4}) is distributed electronically as part
of this paper, with the same projection and layout as the panels of
Figure~\ref{f:canonsnaps}.

Appendix~\ref{a:coulomb} shows that MW-M31 merger times computed from
the semi-analytic orbit integrations are generally in good agreement
with those obtained from $N$-body simulations. In the semi-analytic
Monte-Carlo set, the MW and M31 merge in all of the calculated orbits,
with merger time $t = 5.86^{+1.61}_{-0.72} \Gyr$.  In most of the
Monte-Carlo orbits, the generic properties of the evolution of the MW
and M31 COM are not very different from those in the canonical orbit
discussed in Section~\ref{s:canonical}.  The longest merging time
found was $t = 31.83 \Gyr$, but this is well out in the tail of the
distribution. In general, longer merging times tend to be obtained for
orbits with lower $M_{\rm tot}$ and/or larger $V_{\rm tan}$.

As M31 moves towards the MW, M33 orbits around it. For the M31-M33
system, the first-pericenter time and distance are $t =
0.85^{+0.18}_{-0.13} \Gyr$ and $r = 80.8^{+42.2}_{-31.7}
\kpc$. Pericenter will therefore generally happen within the next
Gyr. The observed velocities imply that the M31-M33 system has a
certain amount of angular momentum, so this rules out a direct hit
between these galaxies at the next pericenter. Technically, a fraction
1.8\% of orbits meet the definition of ``direct hit'' given above, but
these orbits generally have pericenter distances $r > 15
\kpc$. Moreover, orbits with such small pericenters are ruled out by
the argument that on such orbits M33 could not have remained as
symmetric as it is to the present epoch (see
Section~\ref{ss:M33M31}). By the time M31 gets to the MW, M33 is
generally near its second pericenter with M31 (see
Figure~\ref{f:canonsep}).

For the MW-M33 system, the first-pericenter time and distance are $t =
3.70^{+0.74}_{-0.46} \Gyr$ and $r = 176.0^{+239.0}_{-136.9} \kpc$. The
large radial range indicates that M33 does not generally get close to
the MW at its next pericenter. Because of this, M33 on average tends
to reach its pericenter with the MW a little sooner than M31. However,
the distribution of the MW-M33 pericenter distance is extremely broad,
and reaches all the way down to zero. Therefore, in some fraction of
orbits M33 will make a direct hit with the MW, as discussed in more
detail below. This generally happens before M31 reaches its pericenter
with the MW.

We show in Appendix~\ref{a:coulomb} that our semi-analytic orbit
integration method cannot reliably predict the time at which M33 will
merge with either the MW, M31, or their merged remnant. This is due to
the approximate nature of our prescription for the dynamical friction
experienced by M33, which ceases to be valid over timescales $\gtrsim
5 \Gyr$ into the future. Nonetheless, it is possible to draw some
general conclusions based on our $N$-body simulations. The merger time
for M33 will be longer if it settles onto a wider orbit around the
MW-M31 pair. The size of the orbit onto which M33 settles correlates
with the first-pericenter distance between the MW and M33 (compare
Figure~\ref{f:canonsep}, and Figure~\ref{f:orbits} to be discussed
below). Figure~\ref{f:thist}b shows that this distance is smaller for
the canonical model studied in Section~\ref{s:canonical}, than for
68.1\% of Monte-Carlo simulated orbits. Nonetheless, for the canonical
model the M33 merger time was found to be long, since even at the end
of the $10 \Gyr$ simulation, no merger had occurred. This was true
also in a simulation in which M33 settled onto a more radial orbit
around the MW-M31 merger remnant (see
Appendix~\ref{aa:firstM33}). This implies that for most orbits the M33
merger time will generally be considerably longer than $10 \Gyr$. In
contrast, Figure~\ref{f:thist}c shows that the MW and M31 have almost
always merged by then. Therefore, we conclude that the MW and M31 will
generally merge first, with M33 settling onto an orbit around them
that may decay towards a merger later. This result is primarily due to
the mass ratios involved, and not the orbital geometry or angular
momentum: dynamical friction is more efficient at slowing equal-mass
systems than unequal-mass systems.

\subsection{Orbit Classification}
\label{ss:orbitclasses}

The values for the canonical orbit discussed in
Section~\ref{s:canonical} are close to the modes of all distributions
in Figure~\ref{f:thist}. Nonetheless, it is clear from the breadth of
the distribution of MW-M33 pericenter distances that no single orbit
can be a reasonable template for the full variety of possible
outcomes. Nevertheless, we have found that the orbits in the
Monte-Carlo simulations can be broadly classified into the three
categories discussed below, depending on how the distance between M33
and the MW evolves. Table~\ref{t:orbits} lists average properties for
these categories. Figure~\ref{f:orbits} shows example orbits.

{\bf ``M33-hit orbits'':} In 9.3\% of the orbits, M33 ``hits'' the MW
before M31 does. More specifically, these are orbits in which the
distance between M33 and the MW at their first pericenter is less than
25 kpc (which as above, we take as the definition of a direct hit),
while M31 has its own pericenter with the MW either at a later time or
a larger distance. In somewhat less than half of these orbits (4.0\%
of the total), M31 subsequently also makes a direct hit with the MW at
its first pericenter. Orbits in which M33 makes a direct hit with the
MW tend to occur when $M_{\rm M31}$ and $M_{\rm M33}$ have
larger-than-average values, and the M31-M33 relative velocity
$V(M31,M33)$ is smaller than average\footnote{Whether or not
$V(M31,M33)$ is smaller or larger than average depends primarily on
where exactly the M31 and M33 proper motions fall within their
observationally allowed error ellipses.} (see
Table~\ref{t:orbits}). This produces an M31-M33 pair that is more
tightly bound.

{\bf ``Generic M33 orbits'':} In 83.5\% of orbits, M33 does not make a
direct hit with the MW on its first pericenter, but it also does not
move so far from the MW as to ever leave the LG. The canonical orbit
of Section~\ref{s:canonical} is one example of a generic orbit. To
assess whether M33 moves outside the LG, we calculated for each
Monte-Carlo orbit at every time step the distance of M33 from the
barycenter of the MW-M31 pair. If this distance exceeds $0.94 \Mpc$,
the current value of the LG turn-around radius (e.g., Karachentsev
\etal 2002), then M33 was deemed to be outside the (current bounds of)
the LG (this does not take into account any future growth of the LG
and expansion of its turn-around radius). Such an orbit was then
considered not to be a generic orbit.

{\bf ``M33-Ejection orbits'':} In 7.2\% of orbits M33 leaves the LG,
at least temporarily. In this case, M33 can either fall back and merge
with the M31-MW merger remnant much later, or it can become entirely
unbound from the M31-MW system. The fact that M33 can leave the LG
despite being initially bound to M31 (which merges with the MW) is
primarily due to the dynamical friction from the MW on M31. This
causes M31, which is M33's initial center of attraction, to be
dramatically slowed down, while M33 itself keeps moving at the same
velocity.  Orbits in this category tend to occur when $M_{\rm M31}$
and $M_{\rm M33}$ have smaller-than-average values, and the M31-M33
relative velocity $V(M31,M33)$ is larger than average (see
Table~\ref{t:orbits}). This produces an M31-M33 pair that is less
tightly bound.

\subsection{Orbit Examples}
\label{ss:orbits}
 
Figure~\ref{f:orbits} shows the MW-M31-M33 orbital evolution for four
specific sets of initial conditions, to illustrate the categories of
orbits defined in the previous section. For ease of reference we use
the following names for the four models: ``first-M33'', ``canonical'',
``retrograde'' and ``wide-M33''. The canonical model is the same as
discussed in Section~\ref{s:canonical}. The orbital evolution for the
first three models was calculated through $N$-body simulations. The
orbital evolution for the wide-M33 model was calculated with the
semi-analytic orbit-integration method.\footnote{The exclusive goal of
the wide-M33 model was to illustrate the orbit of M33. Since M33 in
this model does not get within $130 \kpc$ of either M31 or the MW,
there was no need for a detailed (and computationally expensive)
$N$-body simulation.} Initial conditions for the four orbits are
listed in Tables~\ref{t:Norbits} and~\ref{t:orbits}, respectively. We
restrict the discussion here to the orbital evolution of the galaxy's
COM. Some selected aspects of the full $N$-body evolution of the
first-M33 and retrograde models are presented in
Appendices~\ref{aa:firstM33} and~\ref{aa:retrograde}. Full movies of
the simulations ({\tt figure\ref{f:rembyM33first}a.mp4} and {\tt
figure\ref{f:retrotails}.mp4}, respectively) are distributed
electronically as part of this paper, with the same projection and
layout as the panels of Figure~\ref{f:canonsnaps}.

The top row in Figure~\ref{f:orbits} shows the first-M33 model, which
provides an example of an M33-hit orbit. The M33 orbit is strongly
curved around M31, sending M33 on a path directed towards the MW. M33
comes within $21.1 \kpc$ of the MW during its first pericenter at
$t=2.91 \Gyr$. M31 is then still $130.2 \kpc$ away, and doesn't reach
its pericenter of $r = 29.9 \kpc$ until $t=3.16 \Gyr$. M33 settles
onto a highly eccentric orbit after the MW and M31 have merged. It
then repeatedly plunges radially back and forth through the center of
the MW-M31 remnant, with slowly decaying apocenters. This particular
model does not have an especially close direct-hit of M33 with the MW
at their first pericenter. A fraction 0.5\% of the orbits in the
Monte-Carlo simulations actually pass within 5 kpc, and a fraction
1.9\% within 10 kpc. However, the first-M33 model does illustrate the
general features of the M31-hit orbit category. For the particular
initial conditions of this orbit, M31 is located at the short end of
its observationally allowed distance range. However, the column of
Table~\ref{t:orbits} that shows the averages for all M33-hit orbits
indicates that this is not a general requirement to end up with a
direct MW-M33 hit.

The second and third rows of Figure~\ref{f:orbits} show examples of
generic orbits, in which M33 does not make a direct hit with the MW at
its first pericenter, and in which M33 does not leave the LG. The
second row shows the canonical model already discussed in
Section~\ref{s:canonical}. In contrast to the first-M33 model in the
top panel, M33 now misses the MW on the negative $Y'$ side at the
first pericenter passage. The third row shows the retrograde model, in
which M33 settles onto a much wider, almost circular orbit around the
MW-M31 merger remnant. The orbital radius is 350--400 kpc and the
period is in excess of 10 Gyr (M33 has not even completed one orbital
revolution by the end of the simulation). The name of this model
derives from the fact that the MW-M31 encounter in this case is
retrograde for both galaxies. However, orbits like this can also arise
with prograde MW-M31 encounters.

The bottom row of Figure~\ref{f:orbits} shows an example M33-out
orbit. In this wide-M33 model, M33 settles onto an orbit that takes it
outside the LG at $t=10.94 \Gyr$, when the LG barycenter distance
reaches $0.94 \Mpc$. It returns back into the LG $8.61 \Gyr$ later.
The maximum distance reached in the meantime is $1.02 \Mpc$ at
$t=15.27 \Gyr$. This apocenter distance is not particularly
extreme. In 3.4\% of the Monte-Carlo simulated orbits M33 actually
reaches further than $1.5\Mpc$ from the LG barycenter, and in 1.0\%
further than $3.0\Mpc$.

Figure~\ref{f:orbits} shows that the different possible categories of
orbits can be viewed as a logical sequence. In the $(X',Y')$ trigalaxy
projection of the COM frame (left panels), from top to bottom, the
initial part of the M33 orbit is characterized by decreasing curvature
towards the COM of the entire system. It is this difference in
curvature that is partly responsible for the different possible
merging outcomes. An important physical quantity that correlates with
this is the current M31-M33 binding energy, which on average decreases
along the sequence.

\subsection{M33-M31 Orbit}
\label{ss:M33M31}

M09 constructed $N$-body models for the M33-M31 interaction to
reproduce features seen in their M33 star count data. They focused on
the past orbit, and the MW was not included. The M31 proper motion was
treated as a free parameter, which was optimized to best reproduce the
generic features of the M33 data.  M09 did not discuss the
quantitative constraints on the M31 proper motion thus obtained, but
they did discuss the properties of the resulting M33-M31 orbit. Their
approach differs from our study in several ways: we use the measured
M31 proper motion, and then focus on the future orbit, including the
MW as well. Nonetheless, it is of interest to examine whether the type
of M33-M31 orbits derived from our analysis are consistent with those
derived by M09 to match the M33 morphology.

M09 found that orbits with previous pericenter distances $r_p \lesssim
40 \kpc$ produced too much distortion of M33 to be consistent with its
overall regular appearance (see also Loeb \etal 2005). Our study
infers only the {\it next} pericenter distance, which exceeds the
previous pericenter distance because of the dynamical friction decay
of the orbit. We find from our orbit calculations (see
e.g. Appendix~\ref{a:coulomb}, Figure~\ref{f:coulomb}b) that the
pericenter decay over one period is typically $\sim 30$\%. The M09
constraint therefore corresponds to $r_p \lesssim 28 \kpc$ at the next
pericenter distance. In our semi-analytic Monte-Carlo calculations
only 2.9\% of orbits have such small pericenters (see the histogram in
Figure~\ref{f:thist}). Therefore, the observed proper motion of M31
from Paper~I is consistent with the overall regular appearance of M33.

M09 detected a warp in the outer stellar distribution of M33,
consistent with the HI morphology. To reproduce this warp in their
simulations, they focused on orbits in which the previous pericenter
distance was not too much larger than $40 \kpc$. They presented
results for one particular orbit that provided a reasonable match to
the generic features of their data. This orbit has a previous
pericenter distance of $53 \kpc$. With the decay rate given above,
this yields a next pericenter distance of $r_p \approx 37 \kpc$. In
Section~\ref{ss:explore} we derived from our Monte-Carlo generated
orbits that the M31-M33 distance at their next pericenter is $r =
80.8^{+42.2}_{-31.7} \kpc$. Therefore, the M09 orbit is $\sim
1.4\sigma$ below the mean of the observationally implied distribution.
This is well within the range of what is plausible, and provides
another successful consistency check on the observed M31 proper
motion.

The next M31-M33 pericenter distance for the canonical model of
Section~\ref{s:canonical} is $r_p = 79.7 \kpc$. This is larger for the
orbit highlighted by M09. The same is true for the other orbits
discussed in Tables~\ref{t:Norbits} and~\ref{t:orbits} (see the values
of $r_p$ listed in the last lines the tables). However, this is not
necessarily a problem, because M09 did not establish an upper limit to
the pericenter distance. They restricted their study to orbits with
previous pericenters $\lesssim 50 \kpc$, and only studied the
evolution in the last $\sim 3.4 \Gyr$. Larger pericenter distances may
well excite acceptable warps, especially if the orbital evolution is
calculated from further back in time, including multiple pericenter
passages.

We have chosen not to restrict the orbits studied here based on the
properties of the M33-M31 orbit, although we do require the pair to be
bound. However, it would have been trivial to further restrict the
initial conditions to those that produce a specific range of
pericenter distances. For example, if we require that the previous
pericenter must have been in the range 40--100 kpc, then this implies
$r = 28$--70 kpc for the next pericenter distance. A fraction 35.3\%
of the semi-analytic Monte-Carlo generated orbits fall in this
range. Of these orbits, 12.5\% can be classified as M33-hit orbits,
85.7\% as generic orbits, and 1.8\% as M33-out orbits. So all of the
different types of orbits of Section~\ref{ss:orbitclasses} are still
present. However, the fraction of M33-out orbits has decreased, since
those orbits have preferentially low M31-M33 binding energies and
large pericenters (see Table~\ref{t:orbits}).

\section{Discussion and Conclusions}
\label{s:conc}

We have studied the future dynamical evolution of the MW-M31-M33
system, using a combination of collisionless $N$-body simulations and
semi-analytic orbit integrations. The initial conditions of this
evolution are well constrained, now that we have determined the M31
transverse motion in Papers~I and~II. The results yield new insights
into the future evolution and merging of the MW-M31 pair. Moreover,
this has been the first MW-M31 study to include detailed models of M33
based on its known transverse motion from water-maser studies.  The
calculations are based on the latest observational and theoretical
insights into the masses and mass distributions of the
galaxies. Monte-Carlo simulations were used to explore the
consequences of varying all relevant initial phase-space and mass
parameters within their observational uncertainties.

We found in Paper~II that the velocity vector of M31 is statistically
consistent with a radial (head-on collision) orbit towards the
MW. This implies that the MW-M31 system is bound, and that the
galaxies will merge. The first pericenter occurs at $t =
3.87^{+0.42}_{-0.32} \Gyr$ from now, at a pericenter distance $r =
31.0^{+38.0}_{-19.8} \kpc$. For the MW, the encounter has 72.2\%
probability of being prograde. For M31, the encounter has 41.4\%
probability of being within $30^{\circ}$ from orthogonal (in terms of
spin-orbital angular momentum alignment). In 41.0\% of Monte-Carlo
orbits M31 makes a direct hit with the MW, defined here as a first
pericenter distance less than 25 kpc. The galaxies merge after $t =
5.86^{+1.61}_{-0.72} \Gyr$.

As M31 moves towards the MW, M33 orbits around it. For the M31-M33
system, the first-pericenter time and distance are $t =
0.85^{+0.18}_{-0.13} \Gyr$ and $r = 80.8^{+42.2}_{-31.7} \kpc$.  The
next pericenter will not be a direct hit, due to the non-zero orbital
angular momentum. The M31-M33 orbit implied by the observed transverse
velocities is broadly consistent with that postulated by M09 to
reproduce tidal deformations in the M31-M33 system. By the time M31
gets to its first pericenter with the MW, M33 is close to its second
pericenter with M31.

For the MW-M33 system, the first-pericenter time and distance are $t =
3.70^{+0.74}_{-0.46} \Gyr$ and $r = 176.0^{+239.0}_{-136.9} \kpc$. The
large range of possible pericenter distances indicates that M33 can
have several different types of orbits with respect to the merging
MW-M31 system.

In 9.3\% of the Monte-Carlo orbits, M33 makes a direct hit with the MW
at its first pericenter, {\it before} M31 reaches pericenter or
collides with the MW. Orbits in this category tend to occur when
$M_{\rm M31}$ and $M_{\rm M33}$ have larger-than-average values, and
the M31-M33 relative velocity $V(M31,M33)$ is smaller than average,
producing an M31-M33 pair that is more tightly bound. In a smaller
fraction of orbits (4.0\%), M31 subsequently {\it also} makes a direct
hit with the MW at its first pericenter.

In 7.2\% of the Monte-Carlo orbits, M33 gets ejected from the LG, at
least temporarily. This is primarily because dynamical friction from
the MW causes M31, which is M33's initial center of attraction, to be
dramatically slowed down while M33 itself keeps moving at the same
velocity. In these orbits, M33 can either fall back and merge with the
M31-MW merger remnant much later, or it can become entirely unbound
from the M31-MW system. Orbits in this category tend to occur when
$M_{\rm M31}$ and $M_{\rm M33}$ have smaller-than-average values, and
the M31-M33 relative velocity $V(M31,M33)$ is larger than average,
producing an M31-M33 pair that is less tightly bound.

In the remaining 83.5\% of Monte-Carlo orbits, M33 does not make a
direct hit with the MW on its first pericenter, and M33 does not move
so far from the MW as to ever leave the LG. In these ``generic''
orbits the MW and M31 will generally merge first, with M33 settling
onto an orbit around them (with a range of possible sizes and
ellipticities) that may decay towards a merger later.

We have explored the orbital evolution of several models through
$N$-body simulations. We have discussed in detail the orbital
evolution and galaxy distortions in a canonical model that produces an
orbit of the generic kind. The initial conditions for this model, as
well as the resulting orbital quantities (e.g., pericenter times,
distances, and merger times), all fall roughly midway in all
observationally allowed ranges. The results of this simulation
therefore provide an ``average'' assessment of what will happen to the
MW, M31, and the Sun in the future.

The radial mass profile of the MW-M31 merger remnant is significantly
more extended than the original profiles of either the MW or M31.  The
profile roughly follows an $R^{1/4}$ profile at radii $R \gtrsim 1
\kpc$, characteristic of elliptical galaxies. This is consistent with
the vast theoretical literature on major mergers of spiral galaxies
(such as the MW and M31) which has found that the remnants of such
mergers resemble elliptical galaxies in many of their properties.

We have analyzed what may happen to the Sun during the evolution of
the MW-M31-M33 system by identifying candidate suns in the canonical
$N$-body model. The Sun could end up near the center of the merger
remnant, but more likely (85.4\% probability) will end up at larger
radius than the current distance from the MW center. There is a 20.1\%
probability that the Sun will at some time in the next $10 \Gyr$ find
itself moving within $10 \kpc$ of M33, but still be dynamically bound
to the MW-M31 merger remnant. The probability that the Sun will become
bound to M33 is much less. While theoretically possible, there were no
candidate suns in this particular simulation that suffered this fate.

The calculations show that the environment of the Sun and the solar
system will be affected by the future MW-M31-M33 orbital
evolution. This includes the Sun's distance from the center of its
host galaxy, its orbit and velocity in the host galaxy, and the local
density of surrounding stars. These changes in environment do not
necessarily imply that the evolution of the Sun and the solar system
themselves would be affected. However, a change in evolution is
certainly possible. For example, the structure of the outer solar
system can be altered by nearby passages of other stars (e.g., Kenyon
\& Bromley 2004). Such passages are generally infrequent, owing to the
collisionless nature of galaxies, and this remains true during galaxy
interactions. Nonetheless, the exact likelihood and severity of such
passages is directly determined by the properties of the Sun's local
environment (Jimenez-Torres \etal 2011), and this will evolve
drastically during the interaction with M31. If a very close passage
were to affect the Earth orbit, it could even affect life (by
relocating the Earth in the solar system relative to the location of
the habitable zone several Gyrs from now).

We have included M33 in our study of the MW-M31 evolution, since it is
the third most massive galaxy of the LG and therefore the satallite
that is most relevant dynamically. It also has a known proper motion,
so that it's future orbit can be calculated. We have not assessed the
future orbital evolution of the many other satellites of the MW, M31,
and the LG. However, it is not impossible that some of the dynamical
features discussed here could apply to other satellites as well. It
would therefore be worthwhile to extend the research presented here
with future calculations and simulations that include more of the
Local Group's satellites. Among other things, this would enable
quantitative study of whether satellites other than M33 may provide a
first direct hit on the MW, whether satellites other than M33 may
leave the LG as a result of the MW-M31 interaction, and whether the
Sun may find itself moving in the future through other satellites than
just M33.

The M31 proper motion measurements discussed in Paper I have allowed
us to obtain new insights into the past, present, and future of the
LG. Paper II focused on the past and the present. It addressed issues
such as the past orbit of the MW-M31 system under the assumption of
the timing argument, and the present-day space velocities and masses
of the galaxies. These have direct relevance for understanding
observational questions related to LG kinematics, cosmology and
stellar archeology. By contrast, Paper III has focused on the future
evolution of the MW-M31-M33 system. This has less direct relevance for
today's observers of the Local Group, since the evolution we calculate
has not yet happened. However, the calculations do have relevance for
other observational questions, e.g., related to the origin of massive
elliptical galaxies. The future evolution we calculate here may
correspond to a specific example of how some of these galaxies and
their satellite systems have evolved to their present state.

The arrival and possible collision of M31 (and possibly M33) with the
MW $\sim 4 \Gyr$ from now is the next major cosmic event affecting the
environment of our Sun and solar system that can be predicted with
some certainty. The other major event that comes to mind, the
exhaustion of the Sun's core of hydrogen fuel, will happen $\sim 2.5
\Gyr$ later (Sackmann, Boothroyd, \& Kraemer 1993). However, as the
Sun's luminosity slowly increases over time, changes in the Earth's
temperature and climate (Kasting 1988; CL08) may well make life on
Earth impossible before M31 and M33 arrive to pay us a visit.

\acknowledgements

The authors are grateful to Mark Fardal, Rachael Beaton, Tom Brown,
and Raja Guhathakurta for contributing to the other papers in this
series, and to the anonymous referee for useful comments and
suggestions. Support for Hubble Space Telescope proposal GO-11684 was
provided by NASA through a grant from STScI, which is operated by
AURA, Inc., under NASA contract NAS 5-26555. The simulations in this
paper were run on the Odyssey cluster supported by the FAS Science
Division Research Computing Group at Harvard University.

\appendix

\section{Coulomb Logarithm}
\label{a:coulomb}

Our semi-analytic orbit-integration methodology uses the Chandrasekhar
formula (Binney \& Tremaine 1987) to describe the dynamical friction
induced by a primary galaxy onto a secondary galaxy. We parameterize
the Coulomb logarithm in this formula as
\begin{equation}
   \log \Lambda = \max [L , \> \log \> (r / C a_s)^{\alpha} ] .
\label{coulomb}
\end{equation}
Here $r$ is the distance of the secondary from the primary, $a_s$ is
the Hernquist profile scale length of the secondary, and $C>0$, $L
\geq 0$, and $\alpha \geq 0$ are constants. This parameterization is
based on the study of Hashimoto \etal (2003). They found that the
Coulomb logarithm must correlate with $r$, to obtain a good
approximation to the $N$-body orbit of a satellite galaxy spiraling
into a more massive primary galaxy. This contrasted with many previous
studies, which often used a constant value of $\log \Lambda$ to
calculate the rate of orbital decay. The use of a floor $L$ for $\log
\Lambda$ is necessary to prevent the dynamical friction deceleration
from becoming an unphysical acceleration for small separations $r < C
a_s$.

The Hashimoto et al.~$N$-body simulations included only a single
component for each galaxy (the dark halo). The satellite was modeled
as a fixed potential, and both galaxies were assumed to have a
constant-density core. For this case, and adopting $L=0$ and $\alpha =
1$, they found a good fit for $C=1.4$. We adopt a more general
parameterization here for several reasons. First, with $L = 0$, there
is no dynamical friction experienced inside $r = C a_s$. This leads to
very slow decay of the satellite once it gets close to the center of
the primary, consistent with what was seen in simulations of Hashimoto
et al. However, this is probably not physical. In real galaxies, the
gravity near the center of the dark halo is dominated by baryons. The
baryons have a higher central density and smaller scale radius than
the dark halo, and therefore provide added dynamical friction when $r
< C a_s$. Moreover, the dynamical response of the satellite adds to
the orbital decay as well. So it is reasonable to consider $L \geq
0$. Also, we allow $C$ and $\alpha$ to differ from the values
advocated by Hashimoto et al. Their values were derived for a
satellite model with an arbitrary constant density core. It is
unlikely that the same values would apply to our cosmologically
motivated models, which have a central dark matter density cusp. The
exact choice of $L$ affects the late-stage evolution of the merger,
whereas the early orbital decay (when the separations are still large)
depends only on $C$ and $\alpha$.

In our orbital calculations we encounter dynamical friction in two
different regimes: dynamical friction between galaxies of roughly
equal mass (MW and M31) and dynamical friction between galaxies of
rather unequal mass (friction on M33 exerted by either the MW or M31,
in both cases corresponding to an $\sim 1$:10 mass ratio); the
dynamical friction of M33 on either MW or M31 is assumed to be
zero. Although we always use the same expression (eq.~[\ref{coulomb}])
for the Coulomb logarithm, it is not obvious that the same values of
the constants $C$, $L$, and $\alpha$ should be used for both cases. We
therefore use different Coulomb logarithm constants
$(C_e,L_e,\alpha_e)$ and $(C_u,L_u,\alpha_u)$ for the (roughly)
equal-mass and unequal-mass case, respectively.

To ``calibrate'' the Coulomb logarithm constants, we used three
$N$-body simulations calculated as described in
Section~\ref{ss:Nmethod}. The first simulation is the canonical
$N$-body model described in Section~\ref{s:canonical}. The second
simulation, which we will refer to as ``calib-1'', has M31 interacting
with the MW in isolation, without M33. Our semi-analytic predictions
for this case depend only on $(C_e,L_e,\alpha_e)$.  The third
simulation, which we will refer to as ``calib-2'', has M33 interacting
with M31 in isolation, without the MW. Our semi-analytic predictions
for this case depend only on $(C_u,L_u,\alpha_u)$. The galaxies in the
calibration simulations had lower masses and higher concentrations
than in the canonical model, as listed in Table~\ref{t:Norbits},

Figure~\ref{f:coulomb} shows the orbital decays $r(t)$ in the $N$-body
simulations (solid curves). We ran many semi-analytic orbit
integrations, varying the Coulomb logarithm constants manually, to
obtain a satisfactory fit to these orbital decays. The fit was judged
by its ability to reproduce the sequence of pericenter and apocenter
distances and times, for the simulations with different galaxy masses
and concentrations. We found that this provided sufficient constraints
to identify a unique set of best-fit parameters. The parameters thus
identified were: $\alpha_e \approx 0.15$, $C_e \approx 0.17$, and $L_e
\approx 0.02$ for the roughly equal-mass case; and $\alpha_u \approx
1.0$, $C_u \approx 1.22$, and $L_u \approx 0$ for the unequal-mass
case. The corresponding orbits obtained from the semi-analytic
calculations are shown as dotted curves in Figure~\ref{f:coulomb}.

The overall agreement between the semi-analytic calculations and the
$N$-body results for the MW-M31 separation in Figure~\ref{f:coulomb}
(red curves) is good. The orbital time scales and separations at
pericenters, apocenters, and merging, are adequately reproduced. The
value $\alpha_e \approx 0.15$ for this roughly equal-mass case is not
far from zero. This indicates that the Coulomb logarithm has only a
mild dependence on radius in this situation.

The value $\alpha_u = 1.0$ {\it inferred} for the unequal-mass case
implies a linear dependence of the impact-parameter ratio $b_{\rm
  max}/b_{\rm min}$ on radius. Interestingly, this is the same as what
was {\it assumed} by Hashimoto \etal (2003) to describe their
unequal-mass simulations. Also, our best-fit $C_u = 1.22$ is very
similar to the value of $1.4$ that they adopted. Nonetheless, we find
that the overall agreement between the semi-analytic orbit
integrations and the $N$-body models in Figure~\ref{f:coulomb} (green
and black curves) is not as good as for the roughly equal-mass MW-M31
case. It is also not as good as found by Hashimoto \etal (2003) for
their unequal-mass simulations. We attribute this to the added
complexities of our simulation compared to those of Hashimoto et al.,
namely multiple galaxy components, cusped halos, and a ``live''
secondary. With these complexities, we find that the long-term
satellite decay is not perfectly fit by the simple formula
eq.~(\ref{coulomb}). In particular, the semi-analytically predicted
decay is too fast at large times. The parameter $L_u$ does not help to
fix this, since increasing it above $L_u \approx 0$ only speeds the
late-stage decay further.

The fact that the orbital decay of M33 is not perfectly reproduced by
the semi-analytical model does not come as a surprise. The decay in
merging and interacting systems is driven to significant extent by
{\it global} responses (e.g., Barnes 1998). These are not well
described by Chandrasekhar's {\it local} dynamical friction
formula. Figure~\ref{f:coulomb} shows that our semi-analytic model is
not adequate to predict the exact time it will take before M33 will
merge with the MW-M31 remnant. To answer this question would require a
suite of $N$-body simulations that follow the satellite merger process
to completion, as in, e.g., Boylan-Kolchin \etal (2008). Nonetheless,
our semi-analytic approach does reproduce the M33 orbital decay
reasonably well for the near-term, $t \lesssim 5 \Gyr$. This is more
than sufficient for the purposes of Section~\ref{s:semicalc}, which
deals with the near-term {\it distributions} of orbital time scales
and properties, and not the long-term details of individual orbits.

The quantitative results for M33's pericenters further illustrate the
adequacy of the semi-analytic calculations for our purposes. For the
first pericenter with M31, the semi-analytic calculations for the
canonical model yield a pericenter distance of $80.3 \kpc$ at $t=0.93
\Gyr$, whereas the actual value from the $N$-body simulations is $79.6
\kpc$ at $t=0.92 \Gyr$. For the first pericenter with the MW, the
semi-analytic calculations for the canonical model yield a pericenter
distance of $74.4 \kpc$ at $t=3.86 \Gyr$, whereas the actual value
from the $N$-body simulations is $97.3 \kpc$ at $t=3.83 \Gyr$. While
the implied error in the M33-MW pericenter distance is $23.1 \kpc$,
this is much smaller than the range of pericenter distances that is
obtained by varying the initial conditions of the orbit
calculations. Figure~\ref{f:thist}b shows that this can yield
pericenter distances ranging from zero to hundreds of kpc. Hence, {\it
uncertainties in the MW-M31-M33 initial conditions (galaxy masses,
positions, and velocities) dominate over uncertainties introduced by
our dynamical friction prescription.}\footnote{As an aside, the
Coulomb logarithm is not the only uncertainty in the amount of
dynamical friction. For example, the dynamical friction at large
separations depends on the uncertain dark halo power-law density
fall-off at large radii. Moreover, this not well resolved in the
$N$-body simulations due to the limited number of dark-halo particles
at large radii. However, any dynamical friction at large separations
is small because of the low densities involved. So here too,
uncertainties in the MW-M31-M33 initial conditions have a much bigger
influence on the exact orbital evolution.}

For the calib-1 simulation, we also show for illustration in
Figure~\ref{f:coulomb}b the prediction for a Kepler orbit (dashed
curve). In this case the MW and M31 were modeled as point masses of
the same total mass as in the $N$-body simulation, and without
dynamical friction. This corresponds to the assumptions on which the
Local Group timing argument is built (see Paper~II). Also, van der
Marel \& Guhathakurta (2008; their figure~3) used Kepler orbits to
constrain the observationally allowed distribution of MW-M31
pericenter distances. Figure~\ref{f:coulomb}b shows that with these
assumptions, obviously, the orbit does not decay after its first
pericenter. Moreover, the Kepler orbit has an earlier pericenter by
0.26 Gyr. This is because the slowing from dynamical friction is
ignored, and because the gravitational attraction is overestimated
when all the mass is assumed to reside at the COM. So while the
semi-analytic predictions obtained in the present paper may not be
perfect, they are certainly a lot more sophisticated than other simple
approaches that have been explored in the context of Local Group
dynamics.

\section{Non-Canonical $N$-body models}
\label{a:Nother}

The initial conditions for the ``first-M33'', ``retrograde'', and
``direct-hit'' models are listed in Table~\ref{t:Norbits}. The
$N$-body evolution of the models was calculated as described in
Section~\ref{ss:Nmethod}. Movies of this evolution are distributed
electronically as part of this paper ({\tt
figure\ref{f:rembyM33first}a.mp4}, {\tt figure\ref{f:retrotails}.mp4},
and {\tt figure\ref{f:direct}c.mp4}, respectively). The same
Galactocentric cartesian $(X,Y)$ projection centered on the MW COM and
the same layout are used as in the panels of
Figure~\ref{f:canonsnaps}. Candidate suns are shown starting at $t=3.0
\Gyr$. The trigalaxy cartesian $(X',Y')$ projection of the orbits
centered on the MW-M31-M33 system COM, as well as the galaxy
separations as function of time, are shown in Figure~\ref{f:orbits}
for the first-M33 and retrograde simulations, and in
Figure~\ref{f:direct} for the direct-hit simulation.

\subsection{The First-M33 Model}
\label{aa:firstM33}

The first-M33 model differs from the canonical model of
Section~\ref{s:canonical} in that M33 passes the MW much more closely
on its first pericenter, $r=21.1 \kpc$ vs.~$97.3 \kpc$. This causes
the subsequent orbit of M33 around the MW-M31 merger remnant to be
more radial. For the first-M33 model the apocenter:pericenter ratio is
13:1 at the end of the simulation ($t=10\Gyr$), whereas it is 2:1 for
the canonical model (see Figure~\ref{f:orbits}). The close passage at
first pericenter in the first-M33 model, and the more radial
subsequent orbit, does not lead to a significantly faster merger of
M33 with the MW-M31 remnant than in the canonical model. In the
first-M33 model, the apocenter distance of the M33 COM at the end of
the simulation is still $63.8 \kpc$. This is slightly, but not
significantly, smaller than the $76.8 \kpc$ for the canonical model.

Figure~\ref{f:rembyM33first} shows the distribution of luminous
particles at the end of the first-M33 simulation. Panel (a) shows the
distribution of all luminous particles, color-coded similarly as in
Figure~\ref{f:canonsnaps}f. Panel (b) shows only the luminous
particles from M33, color coded by local surface density. The latter
can be compared to Figure~\ref{f:rembygal}c for the canonical
model. In both figures, M33 still largely maintains its own identity
in a bound core. However, in the first-M33 model, M33 has shed more
particles into tidal streams and shells that now populate the halo of
the MW-M31 merger remnant. A fraction 64.0\% of the luminous particles
are located further than $17.6 \kpc$ from the M33 COM, compared to
23.5\% for the canonical model. So while the orbit of the most tightly
bound M33 particles is not decaying faster in the first-M33 model
compared to the canonical model, M33 is in fact dissolving
faster. This could be due to the more radial orbit, but the fact that
the galaxy masses in the first-M33 model are higher than in the
canonical model (see Table~\ref{t:Norbits}) may play a role too.

As in the canonical model, no candidate suns became bound to M33
during the first-M33 simulation. However, 100\% of the candidate suns
came within $10 \kpc$ from M33 at some time during the $10 \Gyr$ of
the simulation. This is five times higher than in the canonical
model. However, this is a mere consequence of the radial plunging
orbit of M33 through the MW-M31 merger remnant. It therefore does not
indicate anything of particular interest. For the canonical model it
was more interesting to find candidate suns within $10 \kpc$ from M33,
since M33 itself never came within $23 \kpc$ from the center of the
MW-M31 merger remnant.

\subsection{The Retrograde Model}
\label{aa:retrograde}

M33 stays far from the MW in the retrograde model. Therefore, M33 does
not significantly affect the evolution of the MW-M31 system. However,
this evolution is different than in the canonical model, because the
orbit is now such that both the MW and M31 undergo a retrograde
encounter ($\beta_{\rm MW} = 158.5^{\circ}$ and $\beta_{\rm M31} =
127.6^{\circ}$). For M31 the encounter is still relatively close to
orthogonal, as in the canonical model. However, for the MW the
encounter is now close to maximally retrograde, instead of maximally
prograde (see Figure~\ref{f:prograde}). This impacts the structural
evolution of the MW, and in particular, leads to shorter tidal tails.

Figure~\ref{f:retrotails} shows a snapshot of the simulation at
$t=4.45 \Gyr$, centered on the MW COM, and projected onto the
Galactocentric $(X,Y)$ plane (i.e., the MW disk plane).  This is $0.5
\Gyr$ after the first MW-M31 pericenter, and is close to their first
apocenter. This figure can be directly compared to
Figure~\ref{f:canonsnaps}c for the canonical model. Any MW tidal tails
are less pronounced than in the canonical model. This is likely due to
the retrograde nature of the encounter (Dubinski \etal 1996), but the
fact that the galaxy masses in the retrograde model are higher than in
the canonical model (see Table~\ref{t:Norbits}) may play a role too.

The small MW tidal tails in the retrograde model affect the
distribution of MW particles in the final MW-M31 merger remnant, in
the sense that fewer particles migrate out to very large
radii. Figure~\ref{f:retrosolar} shows the radial distribution of
candidate suns with respect to the center of the MW-M31 remnant, at
the end of the $N$-body simulation ($t = 10 \Gyr$). This can be
compared to Figure~\ref{f:sunhist} for the canonical model. At the end
of the simulation, 18.4\% of the candidate suns reside at $r<
R_{\odot}$ and 81.6\% at $r>R_{\odot}$. A fraction 3.3\% reside at $r
> 50 \kpc$ and a fraction 0.1\% at $r > 100 \kpc$. The fraction of
candidate suns that moves as far out as 50--100 kpc from the MW-M31
merger remnant is three times less than in the canonical model.

As in the canonical model, no candidate suns became bound to M33
during the retrograde simulation. Moreover, no candidate suns came
within $10 \kpc$ from M33 at some time during the $10 \Gyr$ of the
simulation. This is not surprising, given that M33 does not get within
$300 \kpc$ from the MW during the retrograde simulation (see
Figure~\ref{f:orbits}). However, it is interesting in that it
indicates the large range of possible outcomes that is consistent with
the M31 and M33 proper-motion data. By contrast, in the first-M33
model, {\it all} candidate suns came within $10 \kpc$ of M33 at some
time during the $10 \Gyr$ of the simulation. Figure~\ref{f:thist}
shows that the canonical model falls roughly midway in all relevant
MW-M31-M33 orbital parameters. Its predictions with respect to the
fate of the Sun are therefore likely to be close to what one would get
upon averaging over all observationally allowed orbits. It would be
computationally prohibitive to calculate such an average, since it
would require a very large suite of detailed $N$-body simulations.
Nonetheless, it seems reasonable to treat the predictions from the
canonical model with respect to the fate of the Sun as typical for the
overall probabilities one might expect.

\subsection{The Direct-Hit Model}
\label{aa:direct}

\begin{figure}
\epsscale{0.33} \hbox to
\hsize{\plotone{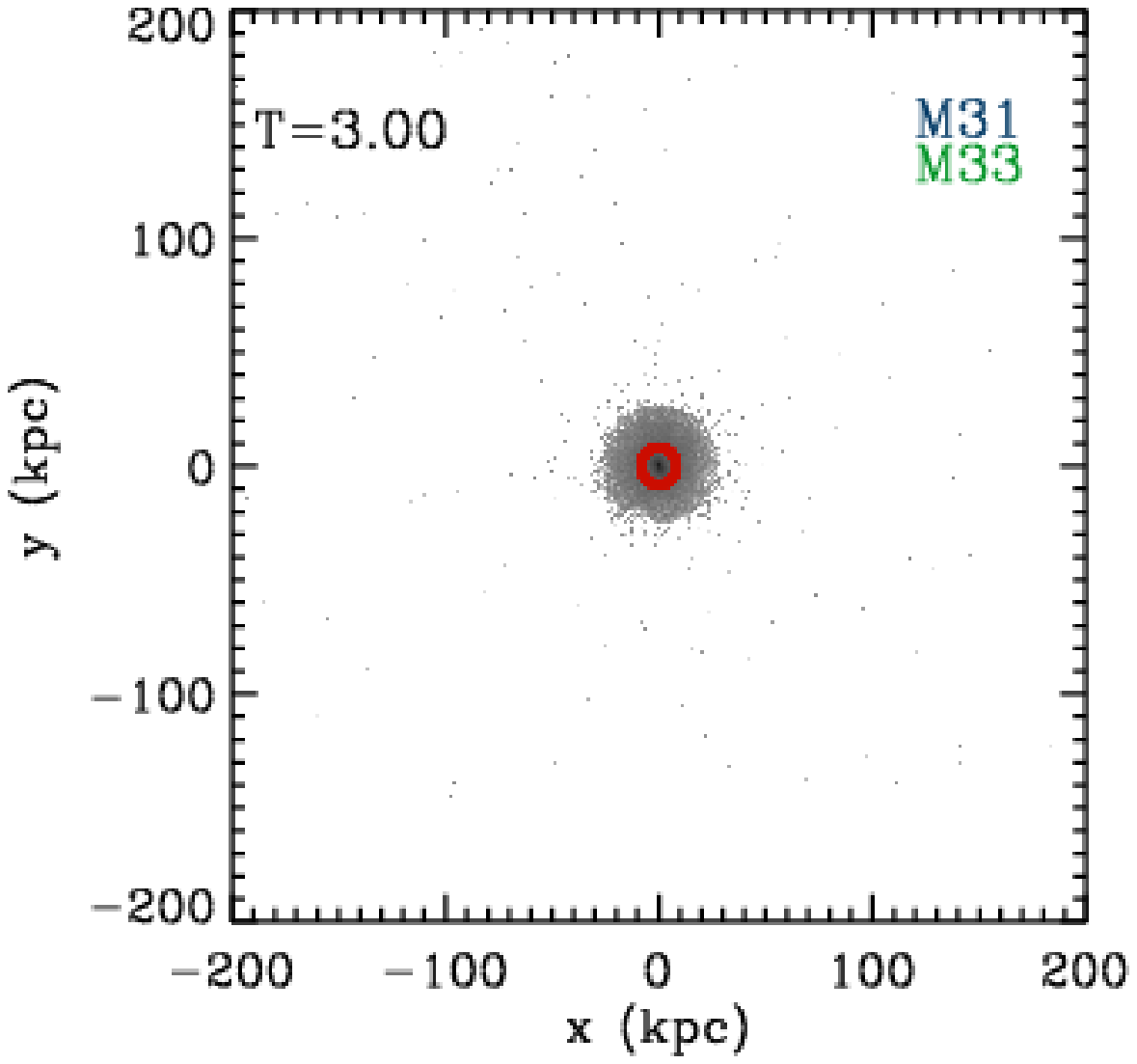}\hfill
       \plotone{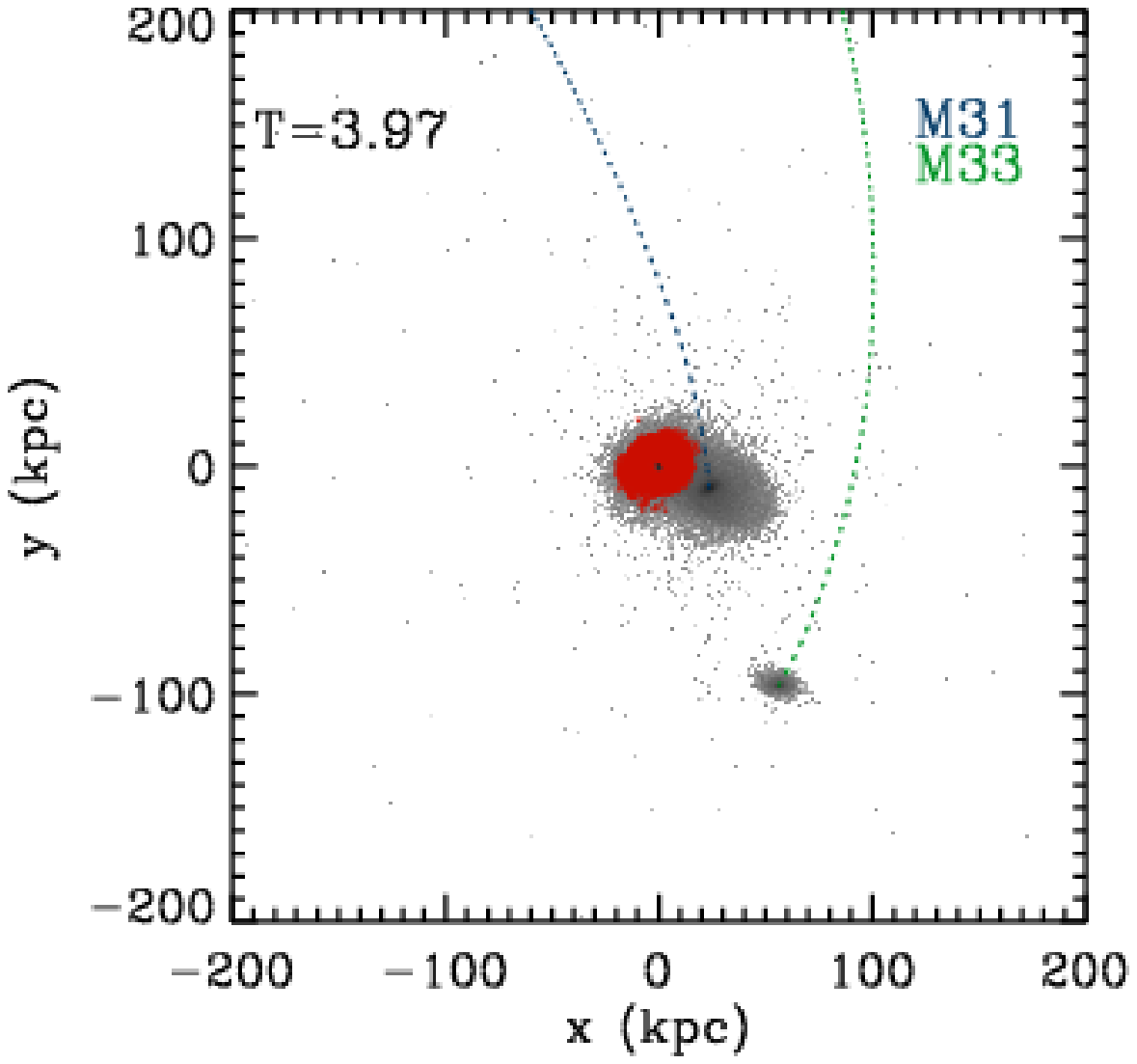}\hfill
       \plotone{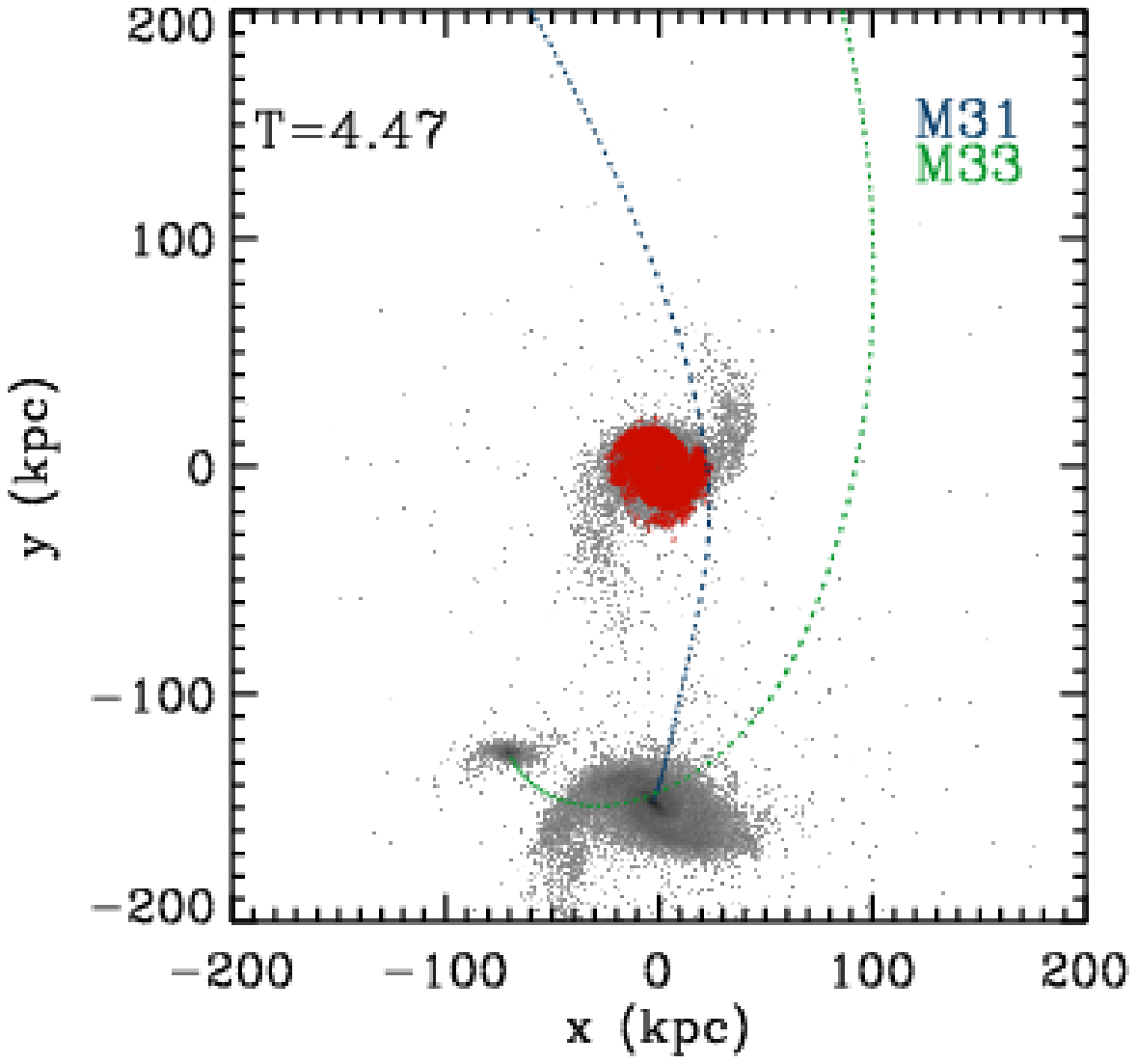}} \hbox to
\hsize{\plotone{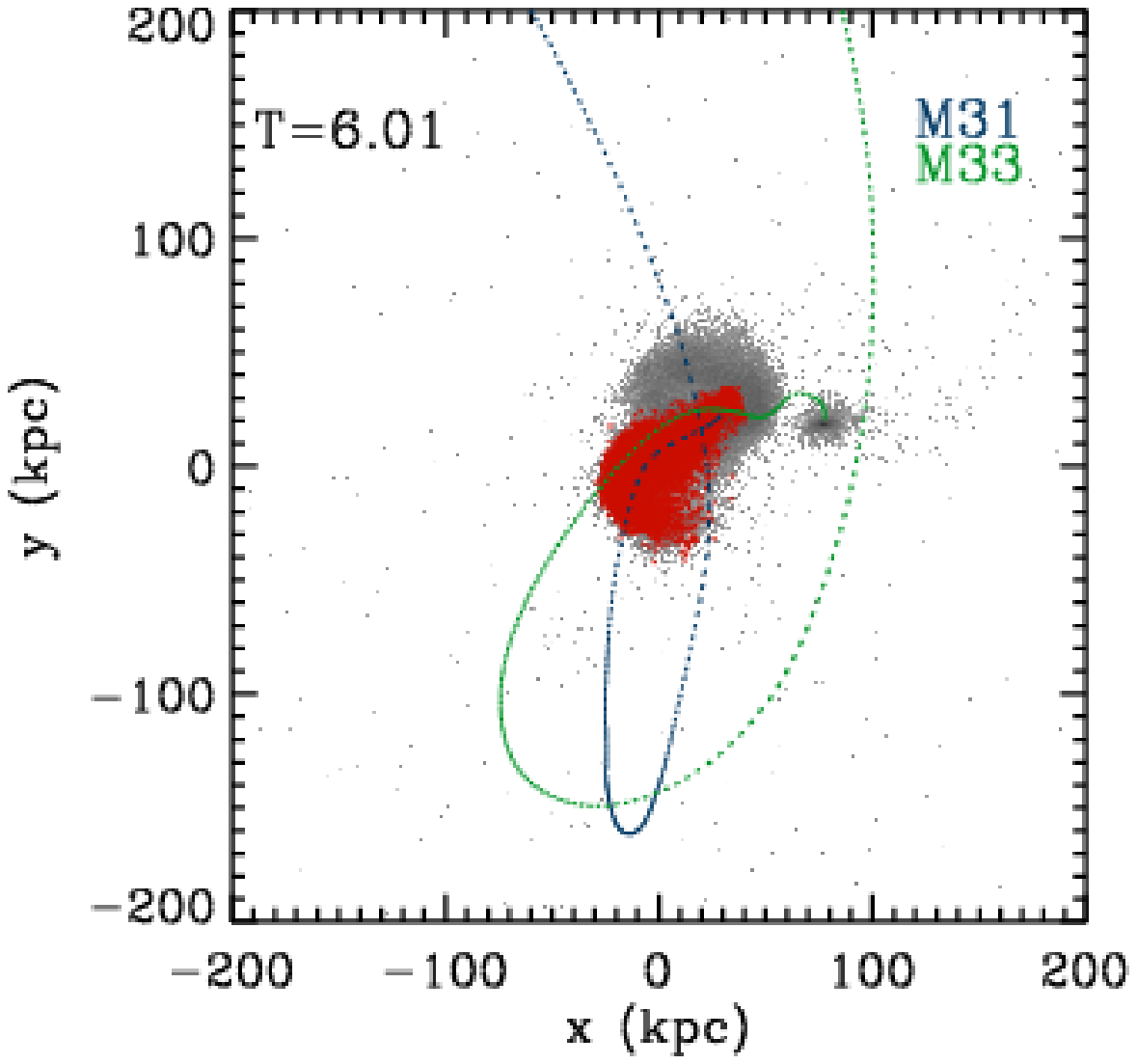}\hfill
       \plotone{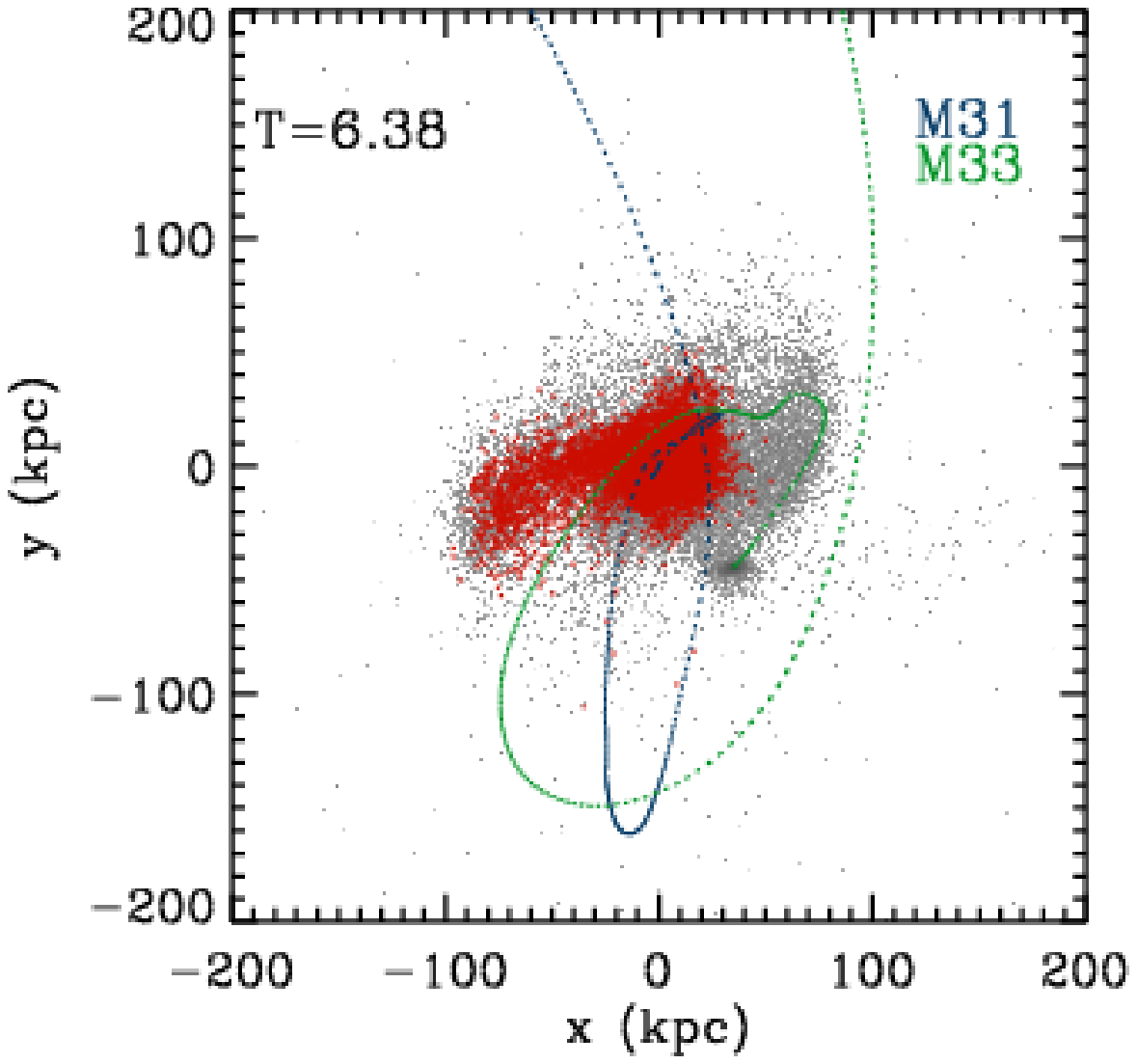}\hfill
       \plotone{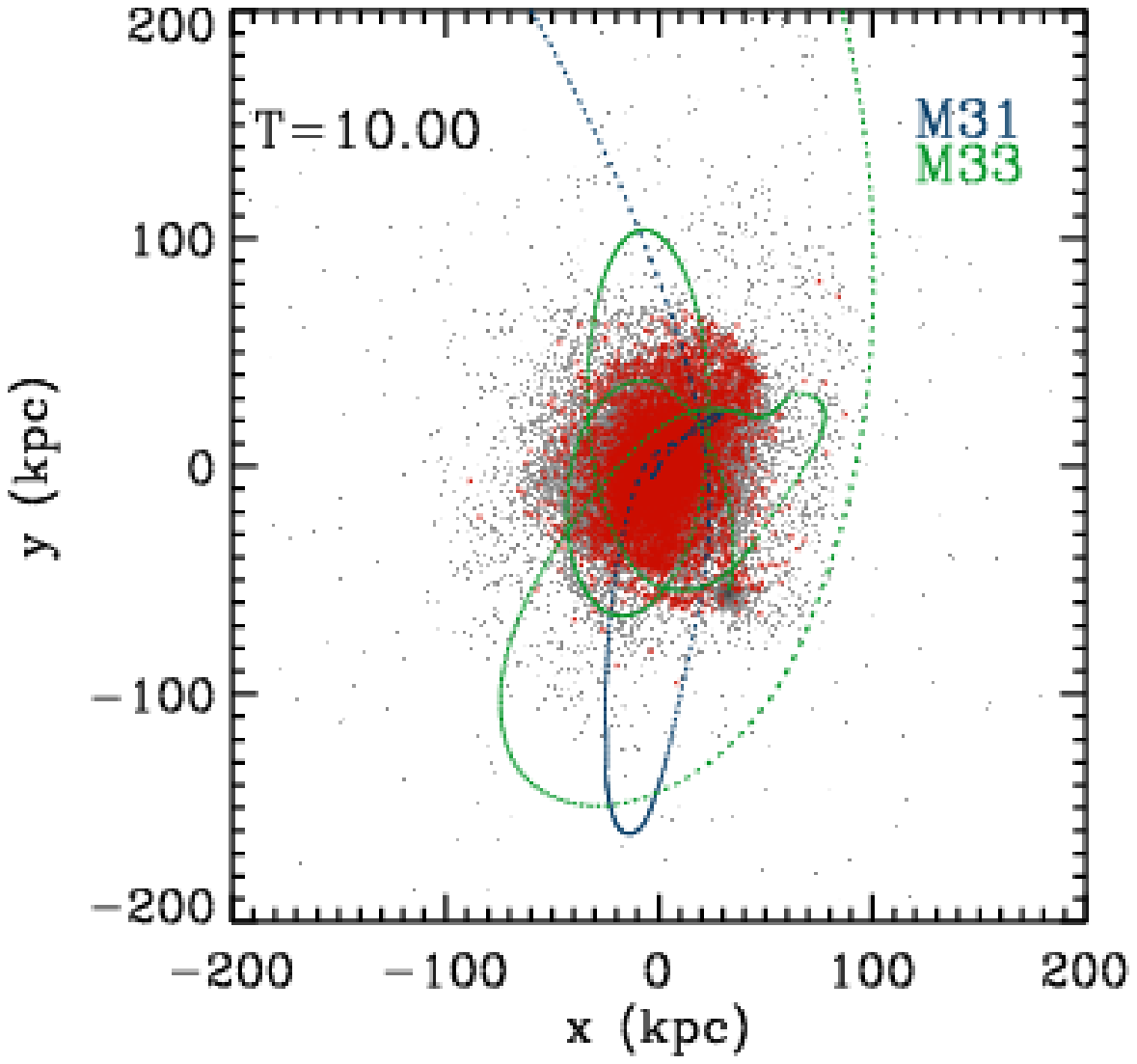}} 
\figcaption{Snapshots of the time
  evolution of the $N$-body simulation for the canonical model,
  centered on the MW COM, and projected onto the Galactocentric
  $(X,Y)$ plane (i.e., the MW disk plane). Only luminous particles are
  shown. Dotted curves in blue (M31) and green (M33) indicate the past orbits
  of the galaxies. The time in Gyr since the current epoch is
  indicated in the top left of each panel. Particles color-coded in
  red are candidate suns, identified as discussed in the text. Panels
  are as follows: (a; top left) Start of the simulation; (b; top
  middle) First MW-M31 pericenter; (c; top right) just before the
  first MW-M31 apocenter; (d; bottom left) second MW-M31 apocenter;
  (e; bottom middle) $\sim 0.1 \Gyr$ after the merger; (f; bottom
  right) end of the simulation.
\label{f:canonsnaps}}
\end{figure}

\begin{figure}
\epsscale{0.33} \hbox to
\hsize{\plotone{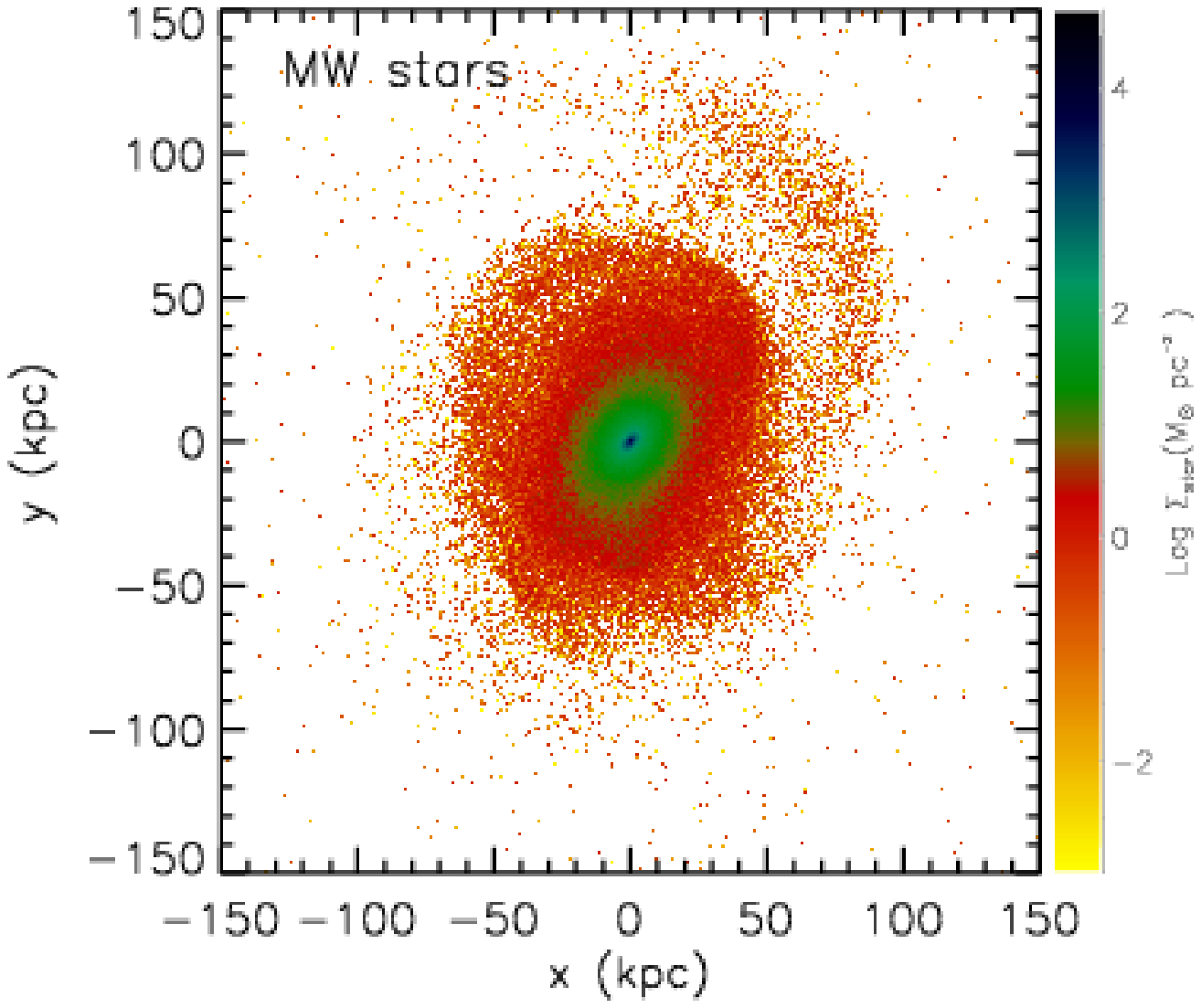}\hfill
       \plotone{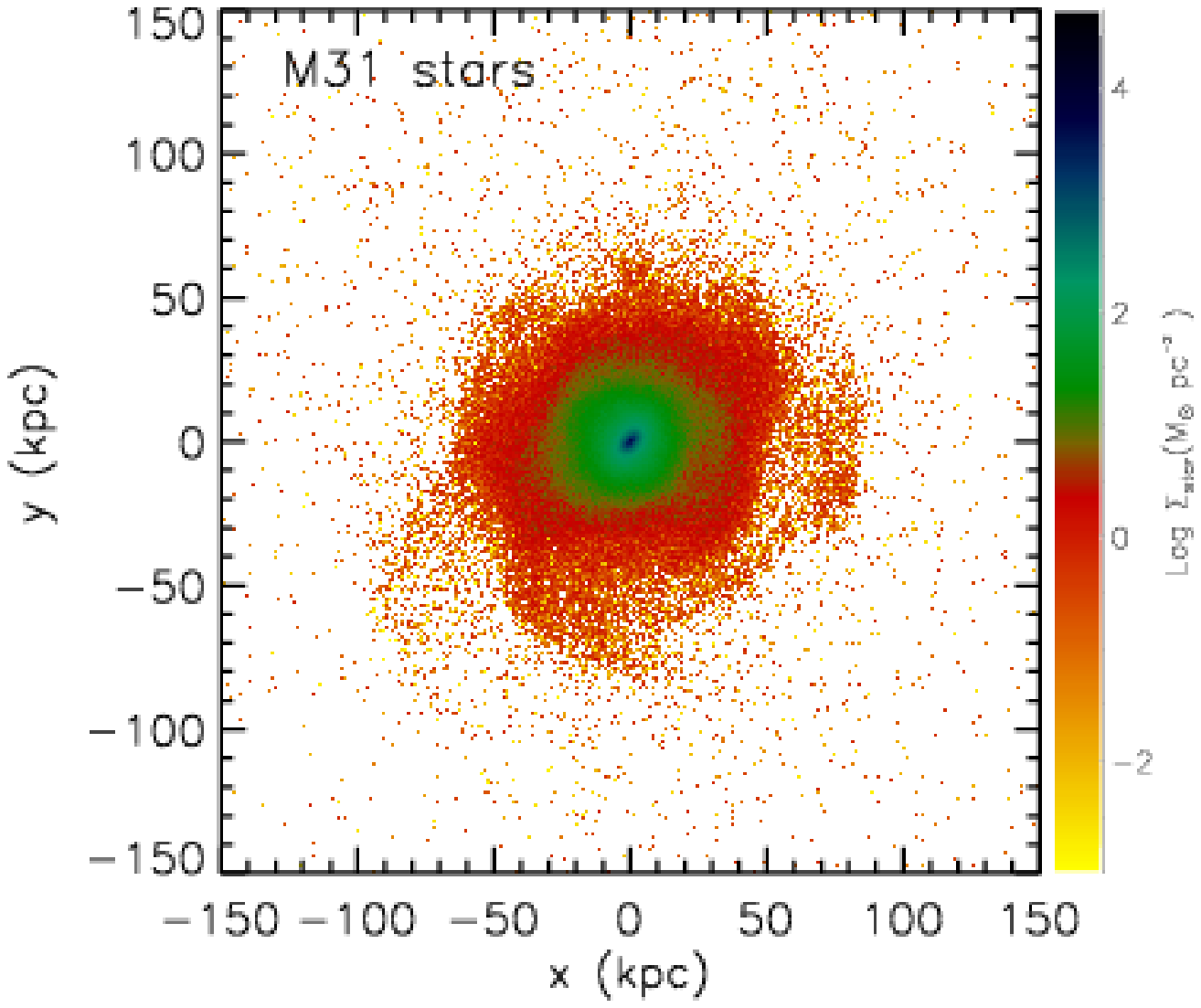}\hfill
       \plotone{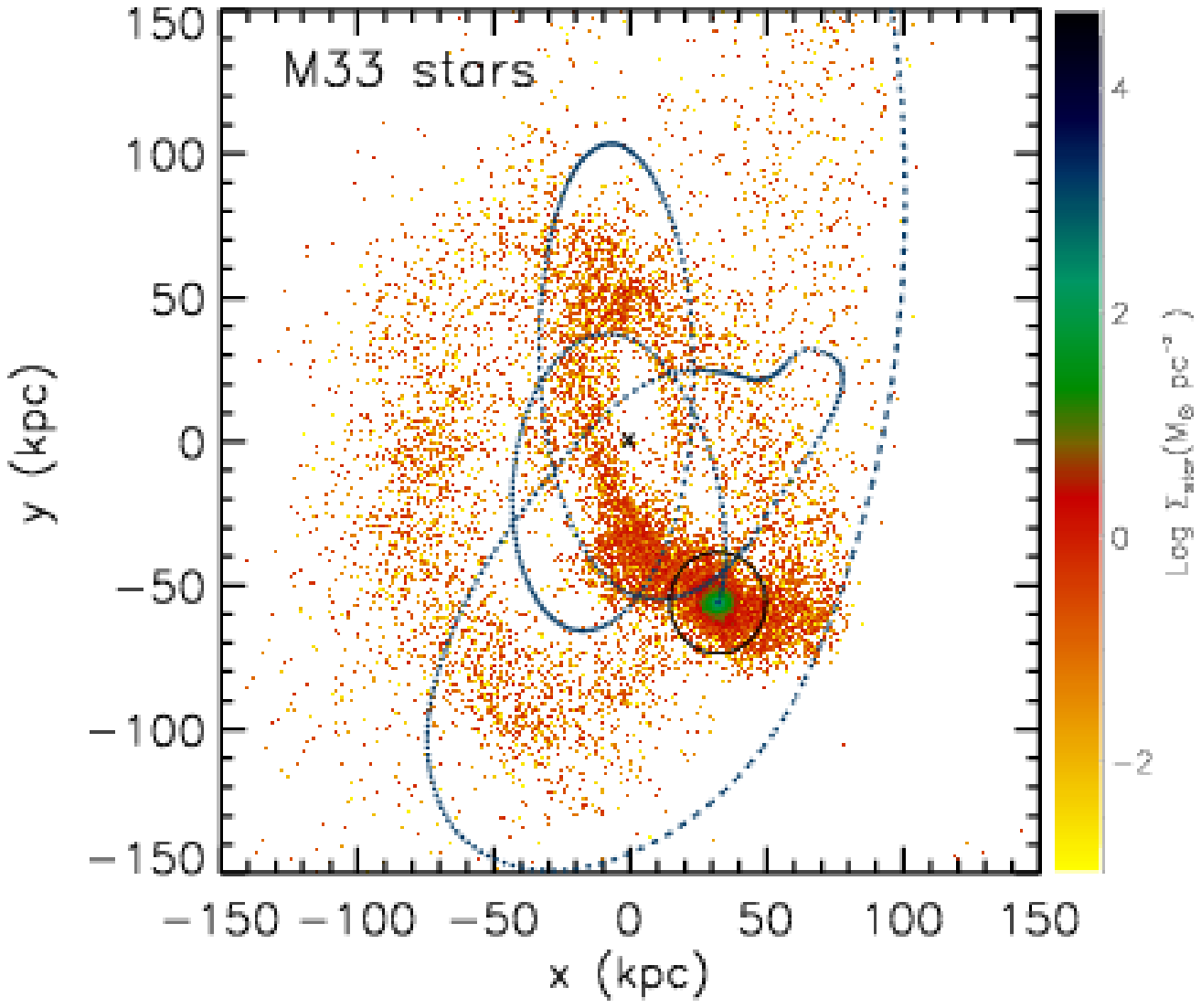}} 
\figcaption{Distribution of
  luminous particles at the end of the $N$-body simulation ($t = 10
  \Gyr$) for the canonical model with, from left to right, particles
  originating in the MW, M31, and M33, respectively. The color scale
  indicates the surface mass density. The COM of each galaxy is at the 
  highest-density position in its particle distribution. 
  For M33, we also indicate the past orbit (dotted
  blue curve), the tidal radius (black circle), and the COM of the MW-M31 
  remnant (black cross). The MW and M31 have
  formed a merged remnant. However, the remnant is not yet fully
  relaxed, since particles originating from the two different galaxies
  still have a somewhat different spatial distribution. M33 maintains
  its own identity, but has lost 23.5\% of its stars into tidal
  streams. These streams do not lie along the location of the orbit.
\label{f:rembygal}}
\end{figure}

\begin{figure}
\epsscale{0.70} \plotone{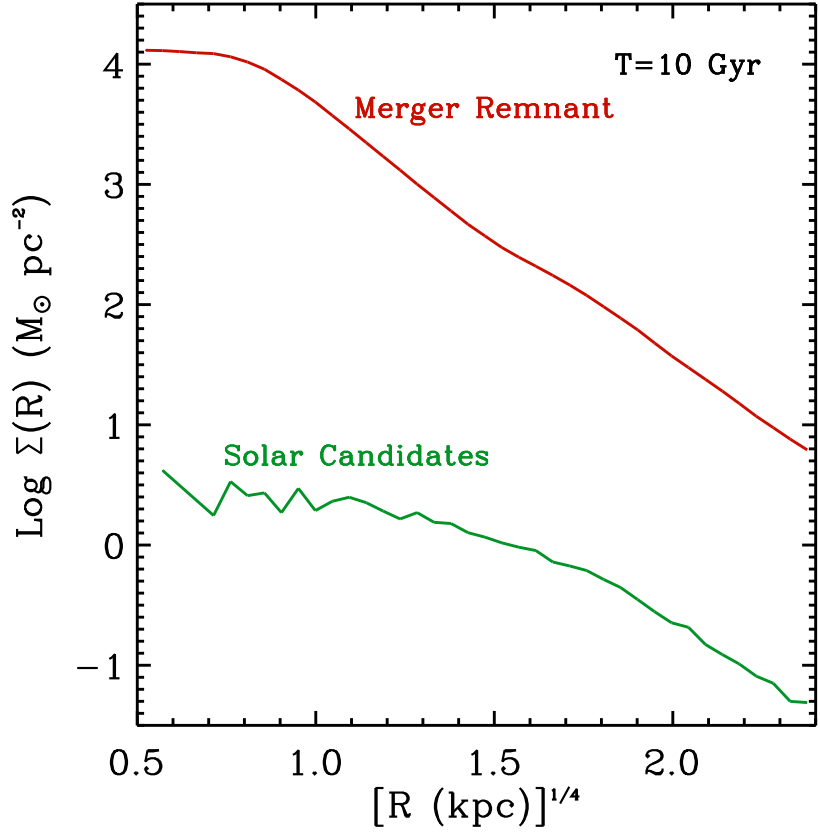}
\figcaption{Projected surface density profile (red) of luminous MW
and M31 particles in the merger remnant at the end of the $N$-body
simulation ($t = 10 \Gyr$) for the canonical model, as function of
$R^{1/4}$ (where $R=(X^2+Y^2)^{1/2}$). The profile is roughly a
straight line for $R \gtrsim 1 \kpc$, indicating it is well
represented by a de Vaucouleurs $R^{1/4}$ law. The profile of
candidate suns (green) is shown as well. It is less centrally
concentrated than the profile of all particles. Hence, a candidate sun
resides on average at a larger radius in the remnant than an average
particle (as is true in the initial MW model as well).
\label{f:remnant}}
\end{figure}

\begin{figure}
\epsscale{0.70} \plotone{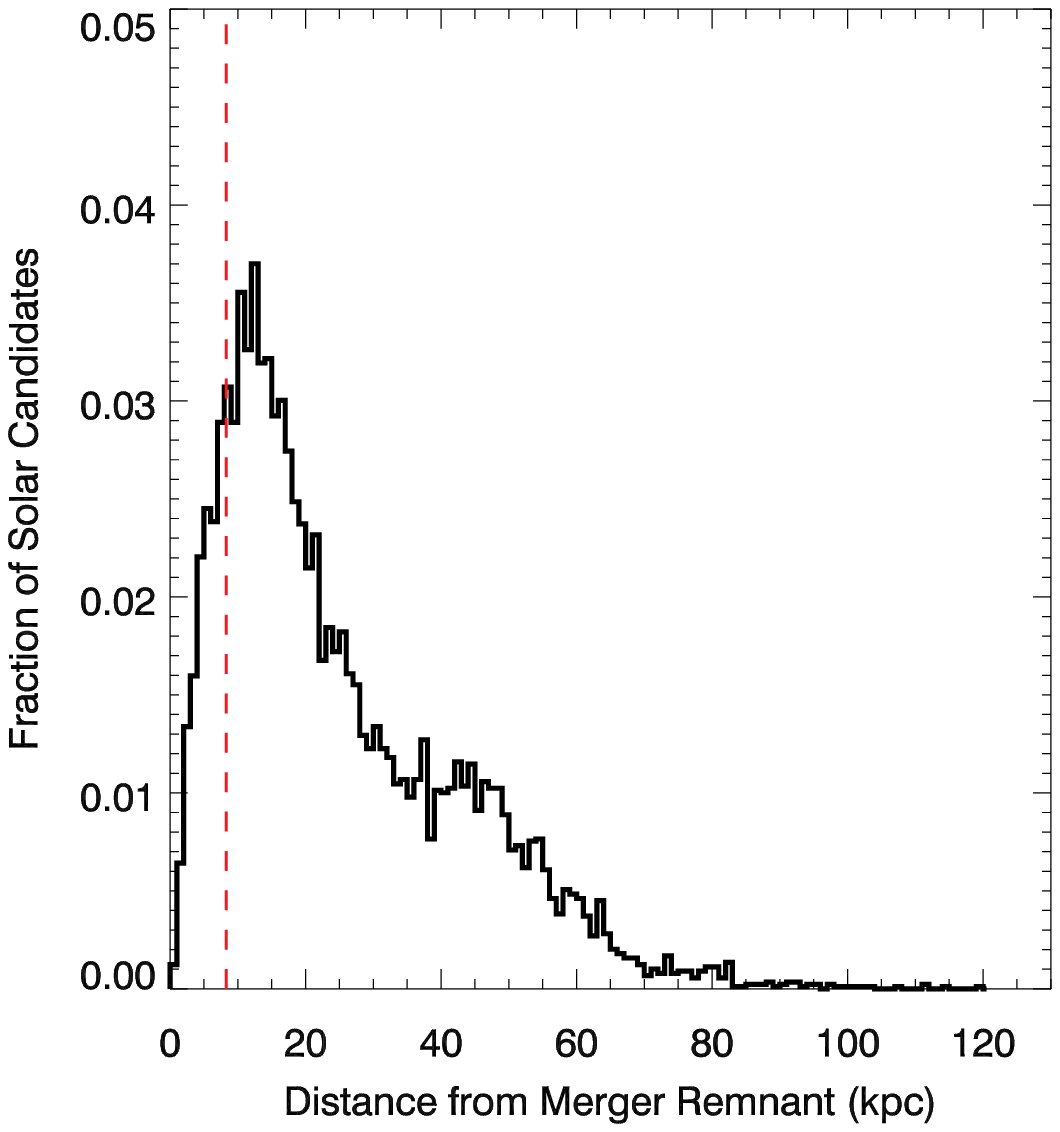} \figcaption{Radial
distribution of candidate suns with respect to the center of the
MW-M31 remnant, at the end of the $N$-body simulation ($t = 10 \Gyr$)
for the canonical model. The red dashed line indicates the current
distance $r \approx R_{\odot} \approx 8.29 \kpc$ of the Sun from the
Galactic Center. All candidate suns start out from that distance. Most
candidate suns (85.4\%) migrate outward during the merger process.
\label{f:sunhist}} 
\end{figure}

\begin{figure} \epsscale{0.6} \plotone{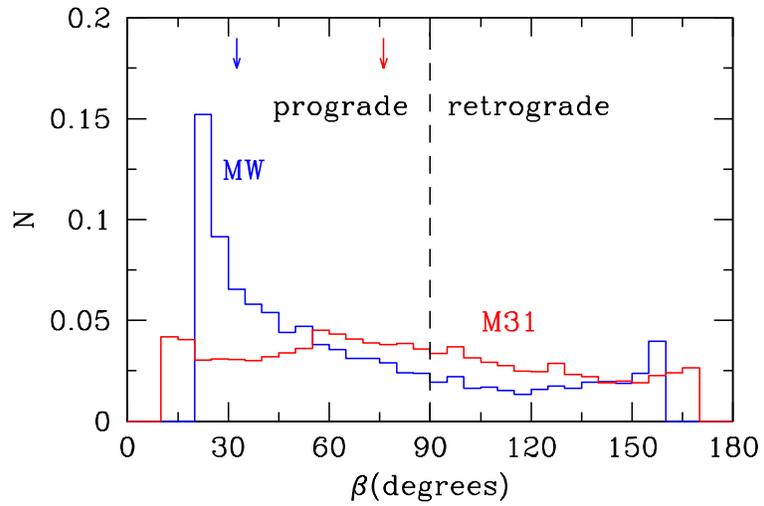}
\figcaption{Histograms of the angle $\beta$ between spin and orbital
angular momentum for the MW (blue) and M31 (red), calculated from the
Monte-Carlo generated initial conditions. The MW is more likely to
undergo a prograde encounter, whereas M31 is more likely to undergo a
nearly orthogonal encounter. Arrows (color-coded
in the same way as the histograms) indicate the values for the
canonical model of Section~\ref{s:canonical}.
\label{f:prograde}} \end{figure}

\begin{figure}
\epsscale{1.0} \plotone{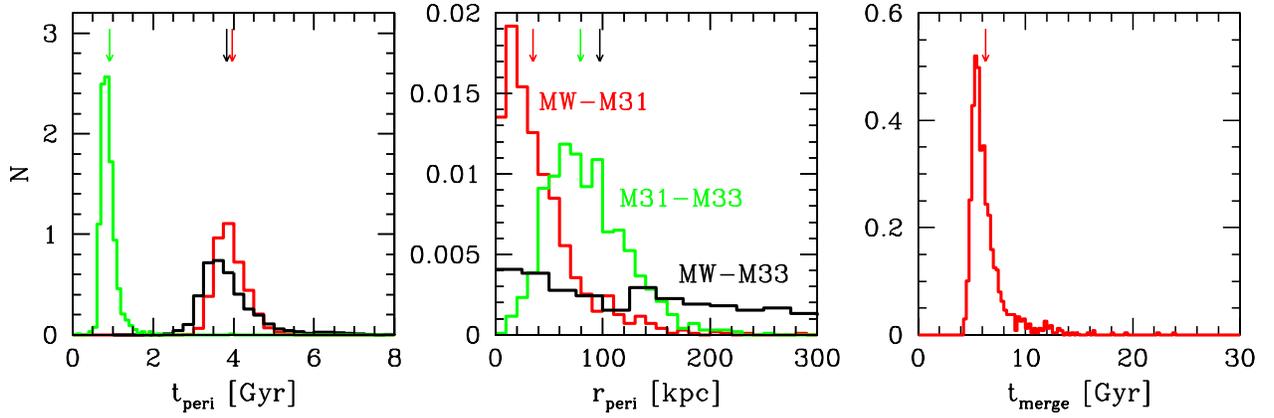} \figcaption{Histograms
  extracted from a Monte-Carlo set of orbits that sample the
  uncertainties in all relevant initial conditions, calculated with
  the semi-analytic orbit integration method. (a) next pericenter time
  $t_p$ for the MW-M31 pair (red), the MW-M33 pair (black), and the
  M31-M33 pair (green). (b) corresponding pericenter distances
  $r_p$. (c) merger time $t_m$. All histograms are normalized to
  unity. Arrows (color-coded in the same way as the
  histograms) indicate the values for the canonical model of
  Section~\ref{s:canonical}. These are in all cases close to the mode or 
  median of the distribution.
\label{f:thist}}
\end{figure}

The initial conditions for the direct-hit $N$-body model were chosen
identical to those for the canonical model, with only one difference:
the initial velocity of the M31 COM was adjusted, within the
observational error bars, to produce a more direct hit of M31 on the
MW. The adopted initial velocity (see Table~\ref{t:Norbits})
corresponds to $V_{\rm tan} = 12.2 \kms$.\footnote{The smallest MW-M31
pericenter is {\it not} obtained for $V_{\rm tan} = 0 \kms$, because
of the perturbing influence of M33 on M31's orbit.} This yields an
MW-M31 pericenter separation of only $3.2 \kpc$ at $t = 3.86\Gyr$,
which is ten times closer than in the canonical model. Comparison of
Figures~\ref{f:direct}a,b to the second row of Figure~\ref{f:orbits}
shows that the orbital evolution is otherwise very similar to that for
the canonical model, although M33 settles onto a somewhat wider orbit
around the MW in the direct-hit model. Figure~\ref{f:direct}c shows a
snapshot of the simulation at $t=4.50 \Gyr$. This is close to the
first apocenter, after the galaxies have already passed through
eachother. This can be compared to Figure~\ref{f:canonsnaps}c for the
canonical model, which has similar layout. A full movie of the
direct-hit simulation is distributed electronically as {\tt
figure\ref{f:direct}c.mp4}.


{}


\clearpage


\begin{deluxetable}{llrrr}
\tablecaption{Galaxy Model Parameters\label{t:galparam}}
\tablehead{Quantity & Unit & Milky Way & \qquad M31 & \qquad M33 \\
(1) & (2) & (3) & (4) & (5)}
\startdata
inclination       & deg            & $\ldots$ & 77.5$^{a,b}$ & 52.9$^d$ \\
PA line of nodes  & deg            & $\ldots$ & 37.5$^{a,b}$ &  4.0$^d$ \\
approaching side  &                & $\ldots$ &   SW$^{a,b}$ &    N$^d$ \\
near side         &                & $\ldots$ &   NW$^{c}$   &    W$^e$ \\
$R_d$             & kpc            &  3.0$^f$ &      5.0$^f$ &  1.3$^g$ \\
$M_b$             & $10^{10}\Msun$ &  1.0$^f$ &      1.9$^f$ &      0.0 \\
$R_b$             & kpc            &      0.6 &          1.0 & $\ldots$ \\
$N_{\rm dark}$    &                &  500,000 &      500,000 &   50,000 \\  
$N_{d*}$          &                &  750,000 &    1,200,000 &   93,000 \\  
$N_{b*}$          &                &  100,000 &      190,000 &        0 \\  
\enddata

\tablecomments{\small Model parameters used in the $N$-body
simulations for the galaxies labeled at the top of columns (3)--(5).
Columns~(1) and~(2) list various quantities and their
units. From top to bottom: inclination; line-of-nodes position angle;
approaching side of the disk; near side of the disk; exponential disk
scale length $R_d$; bulge mass $M_b$; $R^{1/4}$ bulge effective radius
$R_b$; numbers of dark matter particles in the simulations, $N_{\rm
dark}$; numbers of stellar particles in the disk and bulge of the
simulations, $N_{d*}$ and $N_{b*}$. Columns~(3)--(5) list the
quantities for the MW, M31, and M33, respectively. Were relevant,
sources are indicated with superscripts as follows: (a) Chemin \etal
2009; (b) Corbelli \etal (2010); (c) Iye \& Richter (1985); (d)
Corbelli \& Schneider (1997); (e) Corbelli \& Walterbos (2007); (f)
Klypin \etal (2002); (g) Regan \& Vogel (1994). The numbers of
particles pertain to the canonical model discussed in
Section~\ref{s:canonical}. The other simulations listed in
Table~\ref{t:Norbits} used the same mass per particle, yielding
slightly different total particle counts.}
\end{deluxetable}


\begin{deluxetable}{llrrrrrrr}
\tabletypesize{\small}
\tablecaption{$N$-body Initial Conditions\label{t:Norbits}}
\tablehead{Quantity & Unit & 
canonical & retrograde & first-M33 & direct-hit & calib-1 & calib-2 \\
(1) & (2) & (3) & (4) & (5) & (6) & (7) & (8)}
\startdata
$M_{\rm MW,vir} $ & $10^{12} \Msun$  & 1.50   & 1.77   & 2.18  & 1.50   & 1.26     & $\ldots$ \\
$M_{\rm M31,vir}$ & $10^{12} \Msun$  & 1.50   & 1.83   & 1.93  & 1.50   & 1.51     & 1.27 \\
$M_{\rm M33,vir}$ & $10^{12} \Msun$  & 0.150  & 0.079  & 0.249 & 0.150  & $\ldots$ & 0.103 \\
\hline
$c_{\rm MW,vir} $ &                  &  9.56  &  9.45  &  9.30 &  9.56  & 17.02    & $\ldots$ \\
$c_{\rm M31,vir}$ &                  &  9.56  &  9.42  &  9.39 &  9.56  & 16.98    & 17.04 \\
$c_{\rm M33,vir}$ &                  & 11.37  & 11.93  & 10.94 & 11.37  & $\ldots$ & 15.73 \\
\hline
$M_{\rm MW,disk} $ & $10^{11} \Msun$ & 0.75   & 0.70   &  0.70 & 0.75   &  0.50    & $\ldots$ \\
$M_{\rm M31,disk}$ & $10^{11} \Msun$ & 1.20   & 1.20   &  1.20 & 1.20   &  0.70    & 0.70    \\ 
$M_{\rm M33,disk}$ & $10^{11} \Msun$ & 0.09   & 0.07   &  0.09 & 0.09   & $\ldots$ & 0.05 \\
\hline
$D_{\rm M31}$ & kpc                  &  770.0 & 807.7  & 683.1 &  770.0 & 770.0    & 770.0 \\
$D_{\rm M33}$ & kpc                  &  794.0 & 818.6  & 800.0 &  794.0 & $\ldots$ & 794.0 \\
$D(M31,M33)$  & kpc                  &  202.6 & 209.6  & 223.3 &  202.6 & $\ldots$ & 202.6 \\
\hline
$V_{X,M31}$   & km/s                 &   72.6 &   28.6 &  61.6 &   55.4 &  79.0    &  72.6 \\
$V_{Y,M31}$   & km/s                 &  -69.7 & -102.5 & -76.5 &  -90.4 & -71.4    & -69.7 \\
$V_{Z,M31}$   & km/s                 &   50.9 &   40.8 &  58.3 &   28.6 &  40.0    &  50.9 \\
$V_{\rm tan,M31}$ & km/s             &   27.7 &   29.1 &  24.0 &   12.2 &  29.6    &  27.7 \\
\hline
$V_{X,M33}$   & km/s                 &   43.1 &   32.5 &  34.6 &   43.1 & $\ldots$ &  43.1 \\
$V_{Y,M33}$   & km/s                 &  101.3 &  125.7 &  39.6 &  101.3 & $\ldots$ & 101.3 \\
$V_{Z,M33}$   & km/s                 &  138.8 &  175.9 &  80.9 &  138.8 & $\ldots$ & 138.8 \\
$V(M31,M33)$  & km/s                 &  194.5 &  265.2 & 121.3 &  221.4 & $\ldots$ & 194.5 \\
\hline
$\beta_{\rm MW}$  & degrees          &   32.6 &  158.5 &  59.6 &   91.8 & 21.5     & $\ldots$ \\  
$\beta_{\rm M31}$ & degrees          &   76.2 &  127.6 & 108.9 &   42.8 & 51.5     & $\ldots$ \\
\hline
$r_p(M31,M33)$  & kpc                &   79.6 &   63.6 &  70.3 &   74.2 & $\ldots$ & 83.5 \\
\enddata

\tablecomments{\small Initial conditions for the $N$-body simulations
presented in this paper, as labeled at the top of columns
(3)--(8). The canonical model is discussed in
Section~\ref{s:canonical}, and the retrograde, first-M33, and
direct-hit models are discussed in Appendix~\ref{a:Nother}.  The
orbital evolution for these models is shown in the top three rows of
Figure~\ref{f:orbits}, and in Figure~\ref{f:direct}a,b, respectively.
The calib-1 and calib-2 models are discussed in
Appendix~\ref{a:coulomb}.  Columns~(1) and~(2) list various quantities
and their units. From top to bottom: virial masses of the three
galaxies; NFW virial concentrations of the three galaxies; distances
from the Sun to M31 and M33, respectively, and from M31 to M33;
velocity of M31 in the Galactocentric rest frame, and corresponding
tangential velocity component; velocity of M33 in the Galactocentric
rest frame, and corresponding total relative velocity with respect to
M31; spin angles of the MW and M31 with respect to the orbital angular
momentum. The last row lists the distance between M31 and M33 at the
next pericenter. This is not an initial condition, but was inferred
from the calculated dynamical evolution.}
\end{deluxetable}


\begin{deluxetable}{llrrrrrrrr}
\tabletypesize{\small}
\tablecaption{Semi-analytic Initial Conditions\label{t:orbits}}
\tablehead{Quantity & Unit & Observed Range  & wide-M33 &
$\langle$M33-hit$\rangle$ & $\langle$M33-out$\rangle$ \\
(1) & (2) & (3) & (4) & (5) & (6)}
\startdata
$M_{\rm MW,vir} $ & $10^{12} \Msun$  & 0.75--2.25        &   1.91 &
   1.63 $\pm$ 0.42 & 1.62 $\pm$ 0.36 \\
$M_{\rm M31,vir}$ & $10^{12} \Msun$  & 1.54  $\pm$ 0.39  &   1.31 &
   1.77 $\pm$ 0.38 & 1.13 $\pm$ 0.26 \\
$M_{\rm M33,vir}$ & $10^{12} \Msun$  & 0.148 $\pm$ 0.058 &   0.117 &
   0.166 $\pm$ 0.052 & 0.114 $\pm$ 0.033 \\
\hline
$c_{\rm MW,vir} $ &                  &  9.68 $\pm$ 2.29  & 10.14 &
  10.35 $\pm$ 2.38 &  9.02 $\pm$ 2.08 \\
$c_{\rm M31,vir}$ &                  &  9.80 $\pm$ 2.21  &  8.97 & 
   9.50 $\pm$ 1.89 & 10.20 $\pm$ 2.19 \\
$c_{\rm M33,vir}$ &                  & 11.81 $\pm$ 2.60  &  9.01 &  
  11.78 $\pm$ 2.48 & 11.77 $\pm$ 2.66 \\
\hline
$D_{\rm M31}$ & kpc              & 770.0 $\pm$ 40.0  &     737.5  &
   762.4 $\pm$ 31.9    & 747.5 $\pm$ 37.2 \\
$D_{\rm M33}$ & kpc              & 794.0 $\pm$ 23.0  &     806.7  &
   792.0 $\pm$ 22.9    & 804.2 $\pm$ 19.3 \\
$D(M31,M33)$  & kpc              & 207.6 $\pm$ 8.7   &     208.7  &
   205.1 $\pm$  7.0    & 211.8 $\pm$ 10.9 \\
\hline
$V_{X,M31}$   & km/s             &  66.1 $\pm$ 26.7  &     89.0  &
    63.3 $\pm$ 16.3    &  74.4 $\pm$ 22.7 \\
$V_{Y,M31}$   & km/s             & -76.3 $\pm$ 19.0  &    -71.9  &
   -77.3 $\pm$ 12.4    & -76.0 $\pm$ 16.1 \\
$V_{Z,M31}$   & km/s             &  45.1 $\pm$ 26.5  &     24.5  &
    47.0 $\pm$ 14.9    &  38.2 $\pm$ 17.3 \\
$V_{\rm tan,M31}$ & km/s         & $\leq 34.3 (1\sigma)$ & 41.3  &
    23.1 $\pm$ 17.3    &  33.0 $\pm$ 22.6 \\
\hline
$V_{X,M33}$   & km/s             &  43.1 $\pm$ 21.3  &     57.7  &
    47.4 $\pm$ 17.7    &  43.0 $\pm$ 19.7 \\
$V_{Y,M33}$   & km/s             & 101.3 $\pm$ 23.5  &    128.0  &
    85.1 $\pm$ 15.8    & 116.2 $\pm$ 18.6 \\
$V_{Z,M33}$   & km/s             & 138.8 $\pm$ 28.1  &    156.1  &
   116.7 $\pm$ 20.8    & 155.8 $\pm$ 28.1 \\
$V(M31,M33)$  & km/s             & 207.6 $\pm$ 33.9  &    241.1  &
   180.2 $\pm$ 24.2    & 231.1 $\pm$ 32.7 \\
\hline
$\beta_{\rm MW}$  & degrees & 67.5 $\pm$ 43.6 & 31.6 &
63.0 $\pm$ 41.8 & 59.4 $\pm$ 42.6 \\
$\beta_{\rm M31}$ & degrees & 82.1 $\pm$ 43.5 & 29.7 &
90.6 $\pm$ 40.1 & 66.8 $\pm$ 39.2 \\
\hline
$r_p(M31,M33)$    & kpc    & $\gtrsim 28$  & 130.4 & 
71.9 $\pm$ 21.4 & 130.1 $\pm$ 41.2 \\
\enddata

\tablecomments{\small Initial conditions for the semi-analytic orbit 
calculations presented in this paper. The quantities and their units
in columns~(1) and(2) are the same as in
Table~\ref{t:Norbits}. Column~(3) lists the range for each quantity
implied by observations and/or theory, with $1\sigma$ errors, from
Paper II and Section~\ref{ss:explore}. For $M_{\rm MW}$ the full range
is given for which the probability distribution in Figure~4b of
Paper~II is non-zero. The listed ranges were used to draw initial
conditions for the Monte-Carlo simulations of the MW-M31-M33 orbital
evolution, as described in the text. The spin angles $\beta$ follow from
the other initial conditions as described in
Section~\ref{ss:prograde}.  For the M31-M33 pericenter distance in the
last row, column~(3) lists the constraint from
Section~\ref{ss:M33M31}, but this was not used in drawing the
Monte-Carlo initial conditions. Column~(4) lists the initial
conditions for the wide-M33 orbit shown in the bottom row of
Figure~\ref{f:orbits}. Columns~(5) and~(6) list the average and
dispersion for all Monte-Carlo orbits in the ``M33-hit'' and
``M33-out'' categories defined in Section~\ref{ss:orbitclasses}.  The
average and dispersion for the ``generic'' category are not
listed. Since the large majority (83.5\%) of all orbits fall in this
category, their statistics are similar to those listed in column~(3).}
\end{deluxetable}



\clearpage


\clearpage

\clearpage

\clearpage

\clearpage

\clearpage

\clearpage

\clearpage

\clearpage

\clearpage

\clearpage

\begin{figure}
\epsscale{0.75} \plotone{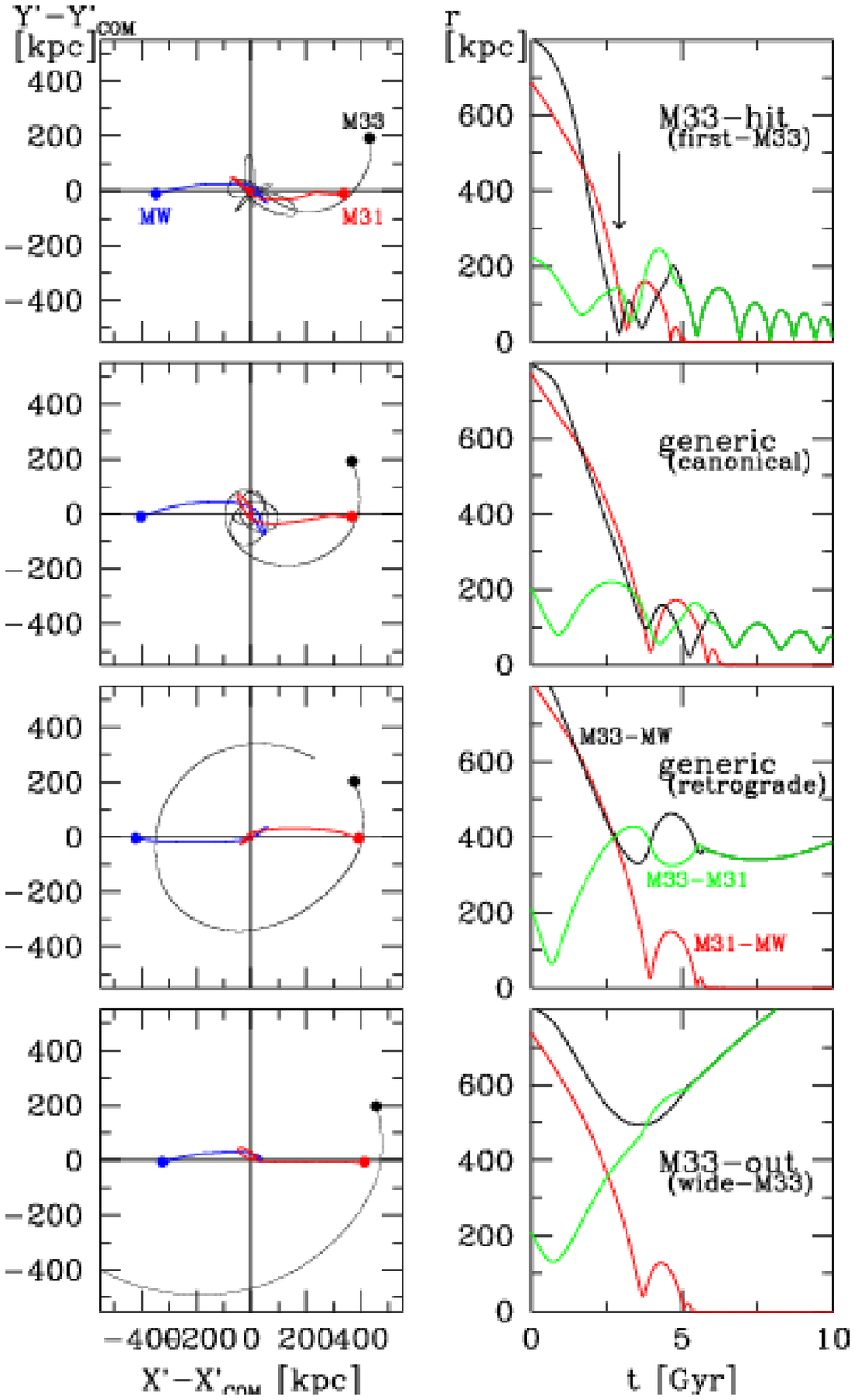}
\end{figure}
\clearpage
\begin{figure}
\figcaption{\small Examples of four types of MW-M31-M33 orbital
evolution, one in each row. The left panel in each row shows the
trigalaxy cartesian (X',Y') projection centered on the system COM, as
in the top row of 
Figure~\ref{f:canonorb}. Positions are shown only for the first $10
\Gyr$. The right panel shows the galaxy separations as function of
time, as in Figures~\ref{f:canonsep}. The initial conditions of the
named orbits are listed in Tables~\ref{t:Norbits}
and~\ref{t:orbits}. The name of each orbit is listed in parentheses in
the right panel, below the name of orbit-category to which it
belongs.(top row) The ``first-M33'' orbit, which is an example of the
class of M33-hit orbits defined in Section~\ref{ss:orbitclasses}. M33
has a close passage with the MW at the time indicated by the arrow,
before M31 encounters the MW.  (second row) The ``canonical'' orbit of
Section~\ref{s:canonical}, which is an example of a generic orbit. M31
and the MW merge, and M33 settles onto an orbit around them that does
not take it outside the LG. (third row) The ``retrograde'' orbit,
which is also an example of a generic orbit. However, in this case M33
settles onto a much wider, almost ciruclar orbit around the MW-M31
merger remnant.  The encounter between the MW and M31 in this orbit is
retrograde for both galaxies. (bottom row) The ``wide-M33'' orbit,
with is an example of the class of M33-out orbits. M33 settles on an
orbit that takes it (at least temporarily) outside the LG. The orbital
evolution for the top three orbits was calculated through $N$-body
simulations, and for the bottom orbit it was calculated with the
semi-analytic orbit-integration method.\label{f:orbits}}
\end{figure}
\clearpage

\begin{figure} \epsscale{1.0} 
\plotone{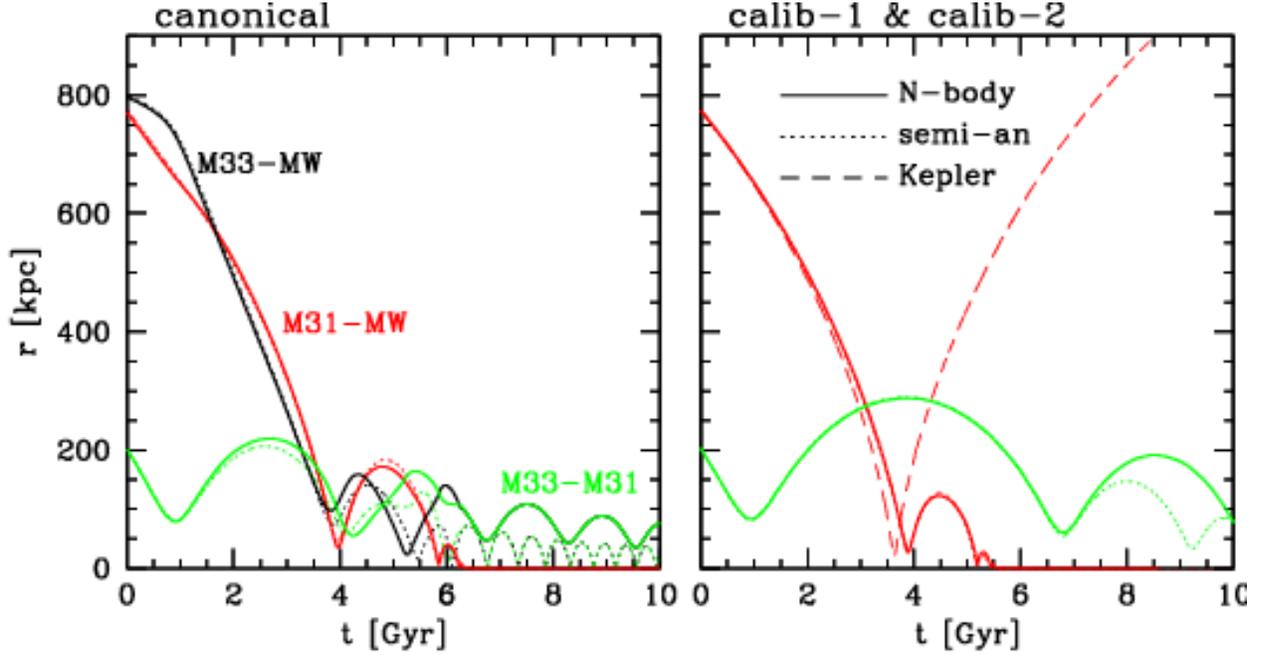} \figcaption{Comparison of the orbital
decay calculated with $N$-body simulations (solid curves) and the
semi-analytic approach with the Columb logarithm discussed in
Appendix~\ref{a:coulomb} (dotted curves). Both panels show the COM
separation vs.~time (counted from the present epoch $t=0$), for the
M31-MW pair (red), the M33-MW pair (black), and the M33-M31 pair
(green), respectively. The left panel is for the canonical model
discussed in Section~\ref{s:canonical}, which includes all three
galaxies mutually interacting. The right panel shows the results of
two different simulations in one and the same panel. The calib-1
simulation (red) includes only the MW and M31, and the calib-2
simulation (green) includes only M31 and M33. The right panel shows
for comparison also the separation for a Kepler orbit of two point
masses of the same mass as the MW and M31 in the calib-1 simulation
(red dashed). Initial conditions of the $N$-body simulations are
listed in Table~\ref{t:Norbits}. Solid and dotted curves overlap in
many places, indicating that the semi-analytic calculations provide a
reasonable description of the $N$-body results. However, the results
for M33 diverge at large times, in the sense that the M33 orbit tends
to decay too fast in the semi-analytic calculations.
\label{f:coulomb}} 
\end{figure} 
\clearpage

\begin{figure} \epsscale{0.33} 
\leftline{\hfill\plotone{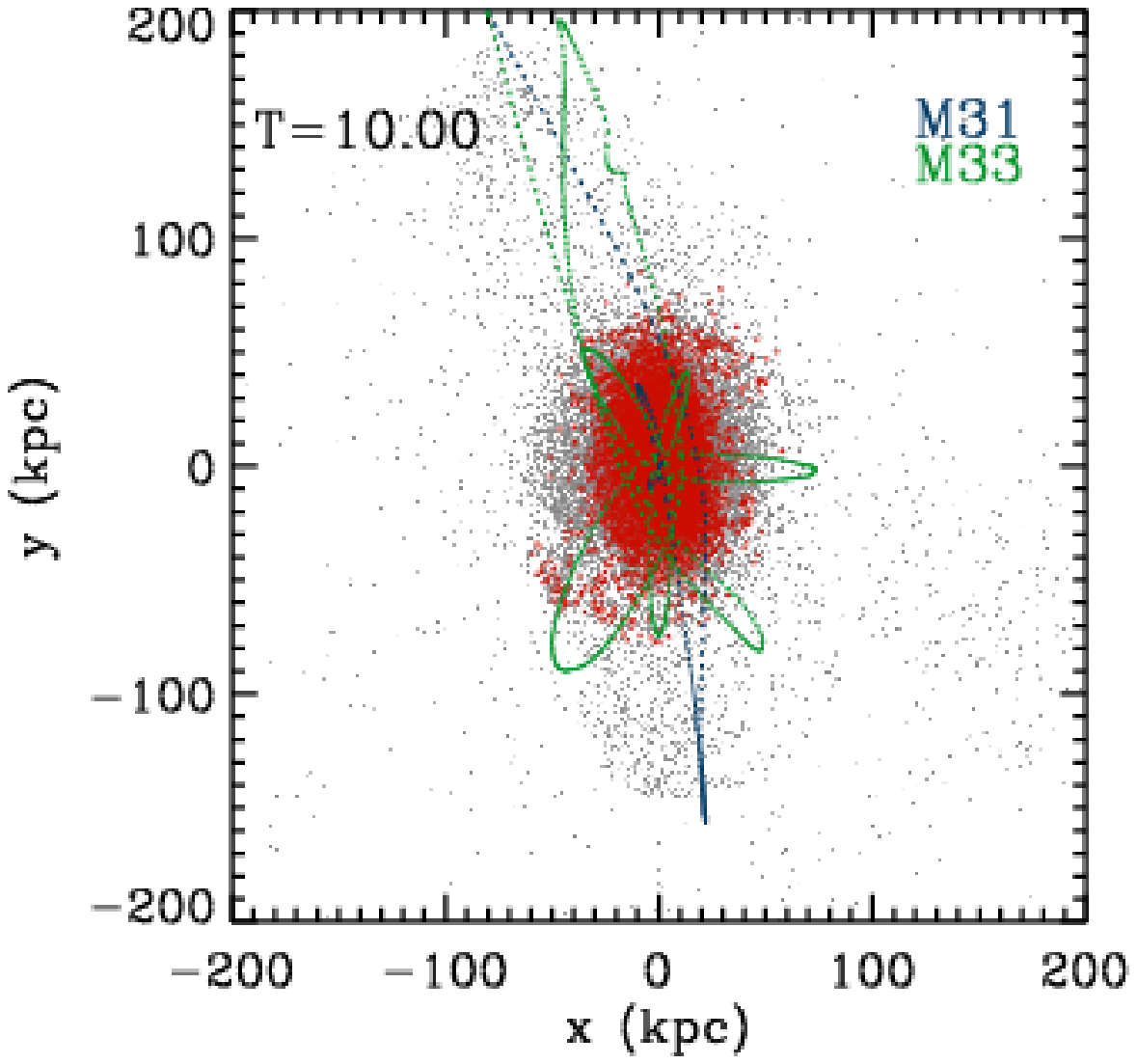}\hfill
\plotone{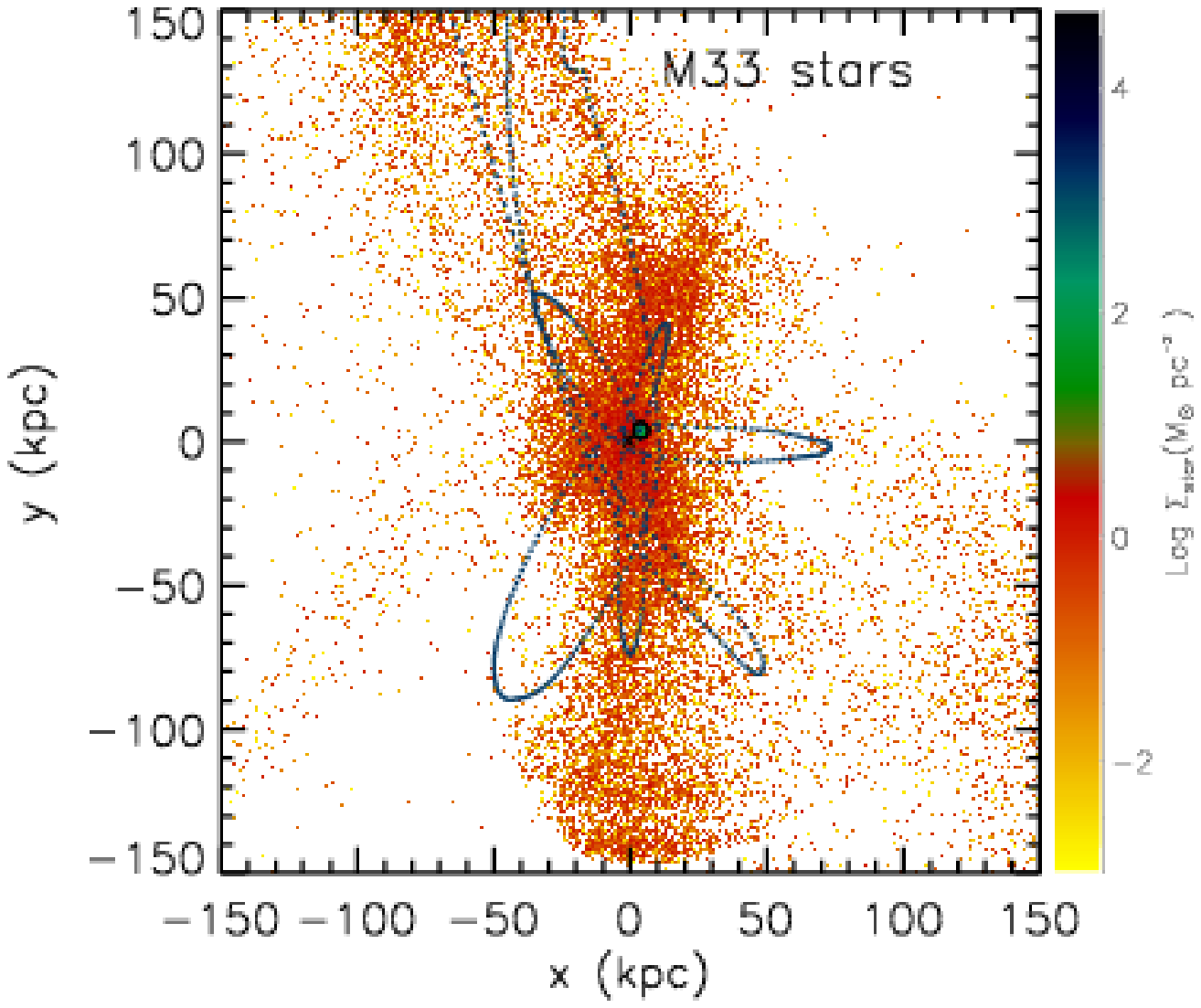}\hfill}
\figcaption{Distribution of luminous
  particles at the end of the $N$-body simulation ($t = 10 \Gyr$) for
  the first-M33 model. The panels show a Galactocentric ($X,Y)$
  projection, at slightly different scales, with the COM of the MW-M31
  merger remnant at the origin. (a, left) distribution of all luminous
  particles, color-coded similarly as in Figure~\ref{f:canonsnaps}f.
  (b, right) distribution of only the luminous particles from M33,
  with the color scale indicating the surface mass density, as in
  Figure~\ref{f:rembygal}. M33 maintains a densely bound core
  (green). This core is near the origin, at its orbital
  pericenter. The past orbit is indicated as a dotted blue curve. M33
  has lost 64.0\% of its luminous particles to distances in excess of
  $17.6 \kpc$. These particles are found in tidal streams and shells
  that now populate the halo of the MW-M31 merger remnant. Their
  location shows some correlation with the past orbit, but not
  accurate alignment for the same reasons as for the canonical model
  (see Section~\ref{ss:remnant}).\label{f:rembyM33first}}
\end{figure} 
\clearpage

\begin{figure} \epsscale{0.33} 
\plotone{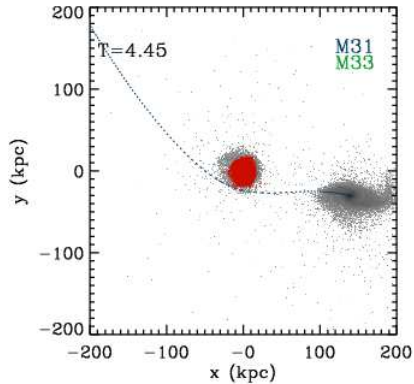}\figcaption{Snapshot of the
$N$-body simulation for the retrograde model, centered on the MW COM,
and projected onto the Galactocentric $(X,Y)$ plane (i.e., the MW disk
plane). Only luminous particles are shown. The blue dotted curve indicates
the past M31 orbit. M33 stays outside the limits of the figure for the
entire simulation (see Figure~\ref{f:orbits}). Particles color-coded
in red are candidate suns, identified as discussed in
Section~\ref{ss:sunfate}. The time of this snapshot is $t=4.45 \Gyr$,
as indicated in the top left. This is just before the first MW-M31
apocenter, and can be compared to Figure~\ref{f:canonsnaps}c for the
canonical model. Due to the retrograde nature of the encounter, the MW
has less well developed tidal tails in the retrograde model than in
the canonical model.\label{f:retrotails}}
\end{figure} 
\clearpage

\begin{figure} \epsscale{0.70} 
\plotone{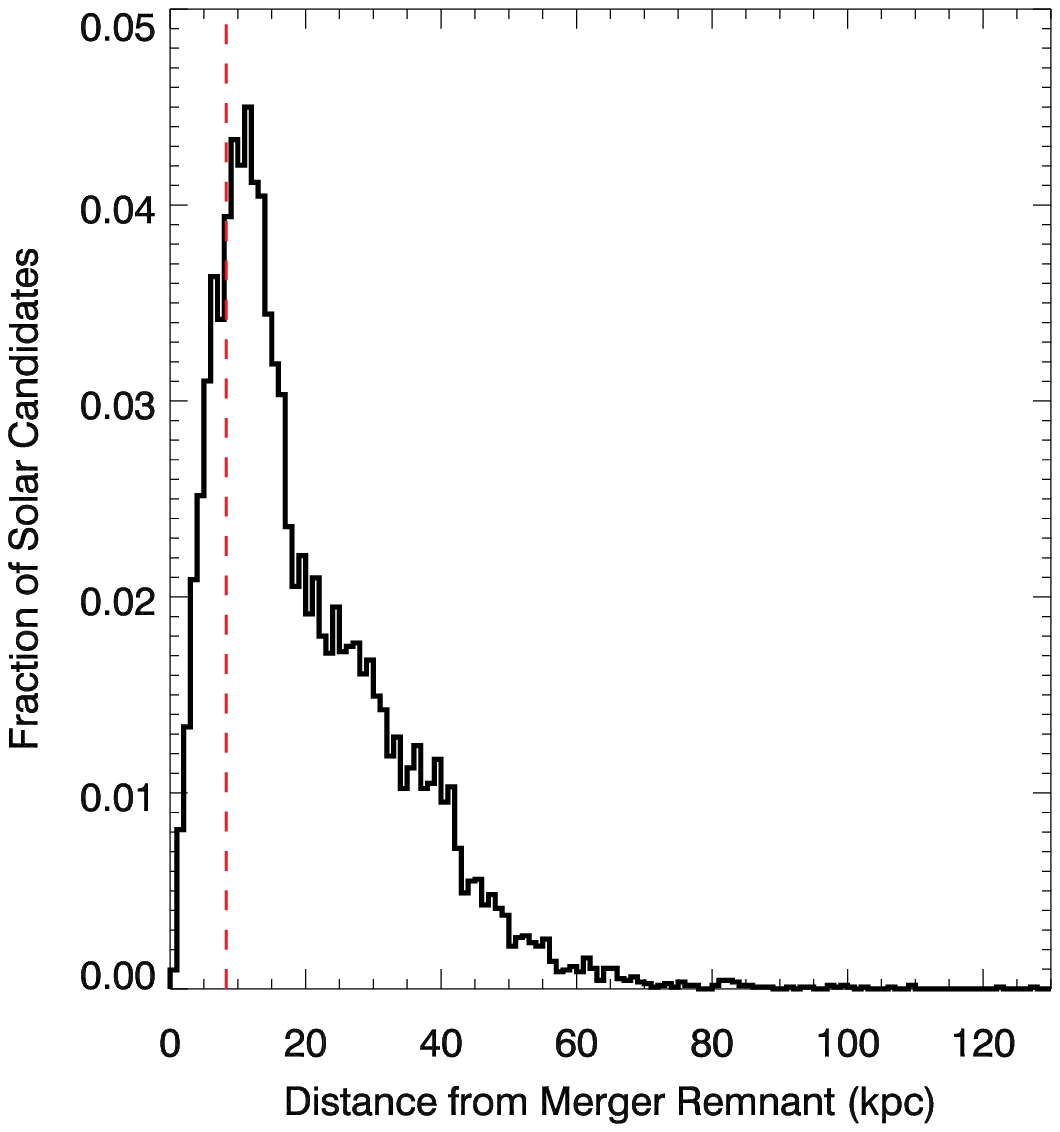}\figcaption{Radial distribution of
  candidate suns with respect to the center of the MW-M31 remnant, at
  the end of the $N$-body simulation ($t = 10 \Gyr$) for the
  retrograde model. The red dashed line indicates the current distance
  $r \approx R_{\odot} \approx 8.29 \kpc$ of the Sun from the Galactic
  Center. All candidate suns start out from that distance. Most
  candidate suns (81.6\%) migrate outward during the merger
  process. However, the outward migration is less on average than in
  the canonical model (Figure~\ref{f:sunhist}).\label{f:retrosolar}}
\end{figure}
\clearpage

\begin{figure} \epsscale{0.70} 
\leftline{\hfill\plotone{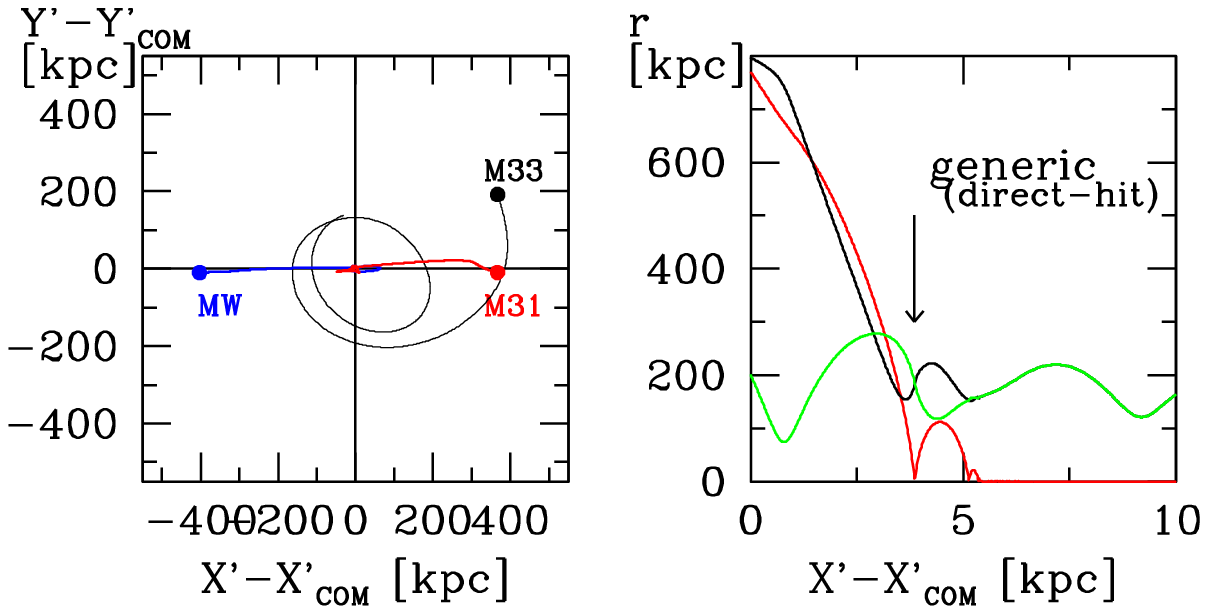}\hfill
\epsscale{0.33}\plotone{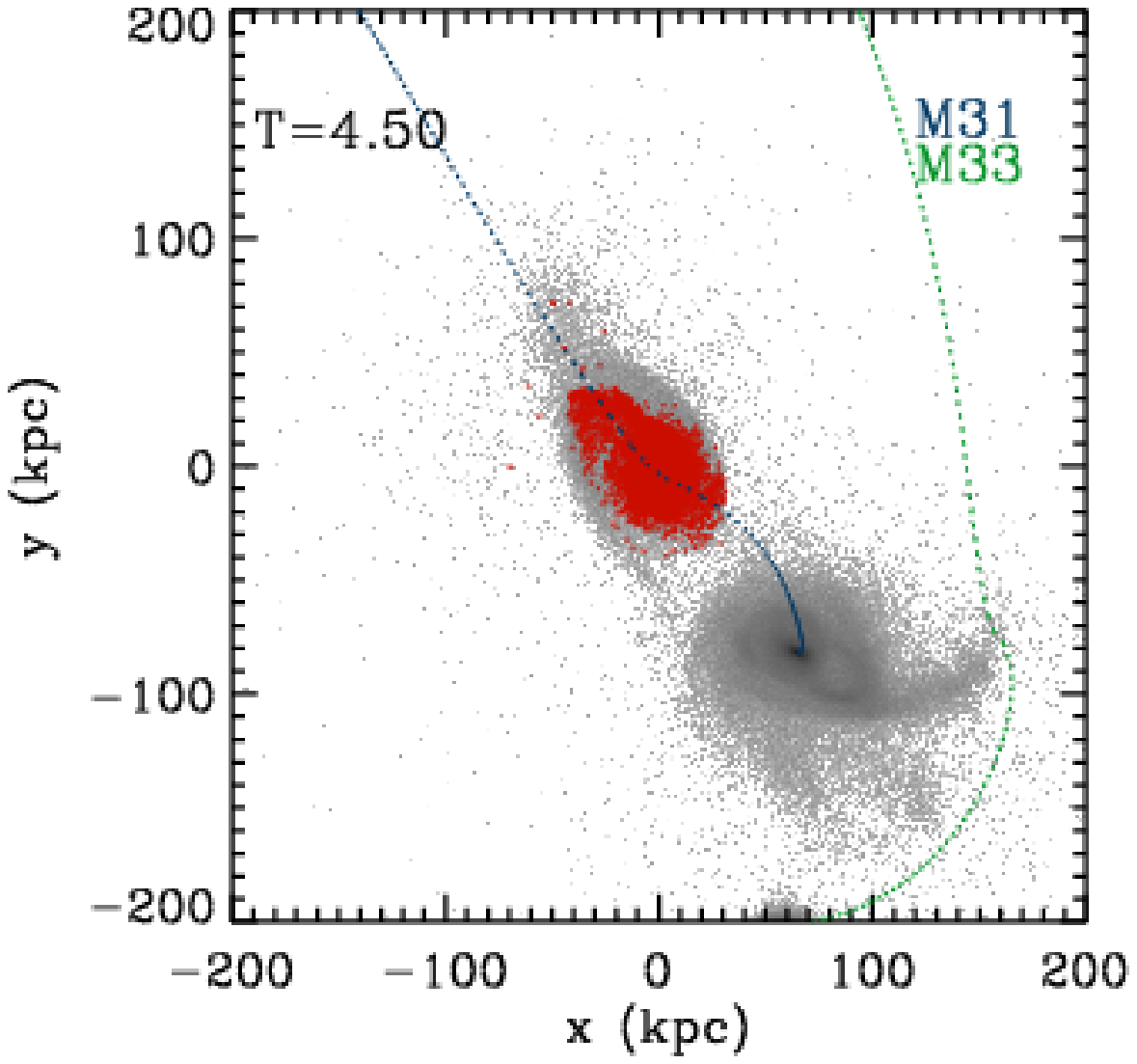}\hfill}
\figcaption{Results of the direct-hit model. (a, left) orbital
evolution shown in the trigalaxy cartesian (X',Y') projection centered
on the system COM, as in the left panels of Figure~\ref{f:orbits}. (b,
middle) galaxy separations as function of time, as in the right panels
of Figure~\ref{f:orbits}. The arrow indicates the pericenter, with a
separation of only $3.2 \kpc$. (c) Snapshot at $t=4.50 \Gyr$, centered
on the MW COM, and projected onto the Galactocentric $(X,Y)$ plane
(i.e., the MW disk plane), as in Figure~\ref{f:canonsnaps}. The blue
dotted curve, marking the past M31 orbit, shows that the MW and M31
have passed straight through eachother. M33 settles onto a wide orbit
around them.\label{f:direct}}
\end{figure}
\clearpage

\end{document}